\documentclass[aps, twocolumn, superscriptaddress, amsmath, amssymb, floatfix, longbibliography, nofootinbib, notitlepage]{revtex4-1}
\usepackage{amsbsy}
\usepackage{amsfonts}
\usepackage{amsmath}
\usepackage{amstext}
\usepackage{amsthm}
\usepackage{amssymb}
\usepackage{graphicx}
\usepackage{epsfig}
\usepackage{epstopdf}
\usepackage{enumerate}
\usepackage{bm}
\usepackage{bbm}
\usepackage{color}
\definecolor{prlblue}{rgb}{0.18,0.19,0.57}
\usepackage{multirow}
\usepackage[colorlinks]{hyperref}
\usepackage{cleveref}
\usepackage{ifthen}
\usepackage{tikz}
\usepackage{multirow}
\usepackage{braket}
\usepackage{comment}
\usepackage[normalem]{ulem}

\usetikzlibrary{patterns.meta} 
\usetikzlibrary{calc}

\newcommand{\tr}{\text{Tr}}

\newcommand{\U}{\mathcal{U}}
\newcommand{\sgn}{\mathrm{sgn}}
\newcommand{\1}{\text{\uppercase\expandafter{\romannumeral1}}}
\newcommand{\2}{\text{\uppercase\expandafter{\romannumeral2}}}
\newcommand{\3}{\text{\uppercase\expandafter{\romannumeral3}}}
\newcommand{\4}{\text{\uppercase\expandafter{\romannumeral4}}}
\newcommand{\5}{\text{\uppercase\expandafter{\romannumeral5}}}
\newcommand{\6}{\text{\uppercase\expandafter{\romannumeral6}}}
\makeatletter

\newsavebox{\@brx}
\newcommand{\llangle}[1][]{\savebox{\@brx}{\(\m@th{#1\langle}\)}%
  \mathopen{\copy\@brx\kern-0.5\wd\@brx\usebox{\@brx}}}
\newcommand{\rrangle}[1][]{\savebox{\@brx}{\(\m@th{#1\rangle}\)}%
  \mathclose{\copy\@brx\kern-0.5\wd\@brx\usebox{\@brx}}}
\makeatother

\def\Z{\mathbb{Z}}
\def\E{\mathcal{E}}

\newcommand{\eqnref}[1]{Eq.\,\eqref{#1}}

\renewcommand{\E}{\mathcal{E}}
\renewcommand{\Z}{\mathbb{Z}}

\newcommand{\be}{\begin{equation}}
\newcommand{\ee}{\end{equation}}

\renewcommand{\tr}{\mathrm{Tr}}

\begin{document}
\graphicspath{{figures/}}

\title{Strong-to-weak spontaneous breaking of 1-form symmetry and \\ intrinsically mixed topological order}

\author{Carolyn Zhang}
\affiliation{Department of Physics, Harvard University, Cambridge, MA 02138, USA}

\author{Yichen Xu}
\affiliation{Department of Physics, Cornell University, Ithaca, New York 14850, USA}

\author{Jian-Hao Zhang}
\affiliation{Department of Physics and Center for Theory of Quantum Matter, University of Colorado, Boulder, Colorado 80309, USA}

\author{Cenke Xu}
\affiliation{Department of Physics, University of California, Santa Barbara, CA 93106}

\author{Zhen Bi}
\affiliation{Department of Physics, The Pennsylvania State University, University Park, Pennsylvania 16802, USA}

\author{Zhu-Xi Luo}
\affiliation{Department of Physics, Harvard University, Cambridge, MA 02138, USA}
\affiliation{School of Physics, Georgia Institute of Technology, Atlanta, Georgia 30332, USA}

\date{\today}

\begin{abstract}

Topological orders in 2+1d are spontaneous symmetry-breaking (SSB) phases of 1-form symmetries in pure states. The notion of symmetry is further enriched in the context of mixed states, where a symmetry can be either ``strong" or ``weak".  
In this work, we apply a Rényi-2 version of the proposed equivalence relation in [Sang, Lessa, Mong, Grover, Wang, \& Hsieh, to appear] on density matrices that is slightly finer than two-way channel connectivity. This equivalence relation distinguishes general 1-form strong-to-weak SSB (SW-SSB) states from phases containing pure states, and therefore labels SW-SSB states as ``intrinsically mixed". According to our equivalence relation, two states are equivalent if and only if they are connected to each other by finite Lindbladian evolution that maintains analytically varying, finite Rényi-2 Markov length. 
We then examine a natural setting for finding such density matrices: disordered ensembles. Specifically, we study the toric code with various types of disorders and show that in each case, the ensemble of ground states corresponding to different disorder realizations form a density matrix with different strong and weak SSB patterns of 1-form symmetries, including SW-SSB. Furthermore we show by perturbative calculations that these disordered ensembles form stable ``phases" in the sense that they exist over a finite parameter range, according to our equivalence relation. 
\end{abstract}
\maketitle

\setcounter{tocdepth}{1} 

{
  \hypersetup{linkcolor=magenta}
  \tableofcontents
}

\section{Introduction}\label{sintro}

Much of modern physics is built on the understanding of symmetries. Recently, even quantum phases that were thought to lie outside of Landau's classification by spontaneous symmetry breaking have been brought into the classic paradigm in a unified way, by generalizing our definition of symmetry \cite{John_review}. In particular, topological orders have anomalous 1-form symmetries which are spontaneously broken, leading to locally indistinguishable degenerate ground states on nontrivial manifolds. When we consider mixed states of matter described by density matrices, the notion of symmetry is further enriched. Unlike pure states, mixed states enjoy two classes of symmetries \cite{de_Groot_2022, MaWangASPT, lee2022symmetry, zhang2024strange, ma2023topological, Luo,  Zhang_2023, ChenGrover2024, chen2024symmetryenforced, sang2023mixed, sang2024Markov, hsin2023anomalies, li2023exact, chen2024unconventional, ma2024symmetry, xue2024tensor, guo2024locally, lessa2024mixedstate, wang2024intrinsic, wang2024anomaly, chirame2024stable, zhang2024quantum, guo2024design, lessa2024strongtoweak, sala2024spontaneous, xu2024average, huang2024hydro, li2023exact}: for $U$ a unitary representation of the 0-form symmetry,\footnote{Symmetries can also be non-invertible. In this work, we will only consider invertible linear symmetries, except for a natural antilinear symmetry that arises in the context of Choi states (explained later in the text).} $U\rho\propto \rho$ indicates that $U$ is a strong symmetry of $\rho$, while $U\rho U^{-1}=\rho $ means that $U$ is a weak symmetry of $\rho$. Recently there has been much progress on different patterns of spontaneous symmetry breaking (SSB) that arise in the context of mixed states, involving both strong and weak 0-form symmetries \cite{lee2, ma2023topological, lessa2024strongtoweak, sala2024spontaneous, xu2024average, huang2024hydro, gu2024spontaneous, moharramipour2024symmetry, su2024emergent, Lu_2020, li2024replica, kuno2024strong}. 

The above definitions of strong and weak symmetries apply to higher-form symmetries as well.\footnote{One has to be careful about how to define emergent symmetries, because for higher-form symmetries we generally allow sufficiently weak explicit breaking of the symmetry. We discuss this in detail in Sec.~\ref{semergent}.} It is thus natural to consider the possibility mixed state topological order, from the perspective of 
SSB patterns of strong and weak 1-form symmetries. Mixed states with higher form symmetries have mostly been studied in the context of topological orders under decoherence channels. Refs. \cite{bao2023mixedstate,lee2,fan2024diagnostics, Zou_2023, wang2024intrinsic, wang2024analog, hauser2024information, kikuchi2024anyon,PRXQuantum.5.040313} studied mixed-state topological order obtained by corrupting pure-state topological orders with  noises. For small enough decoherence strength $p$ (such that the errors can still be corrected and the logical state on a torus can be recovered in the thermodynamic limit), the mixed state retains many of the properties of the pure state. In Ref. \cite{li2024replica}, a state with $p<p_c$ (where $p_c$ is the critical decoherence strength) is called a  
quantum memory, while one with $p>p_c$ is called a classical memory. Refs. \cite{zini2021,wang2024intrinsic,sohal2024noisy, ellison2024classification} showed that the spontaneously broken strong 1-form symmetries for mixed states only need to form premodular tensor categories rather than modular tensor categories. Premodular tensor categories describe anyon theories where the anyons do not need to satisfy braiding non-degeneracy, and therefore may have degenerate $S$-matrices. 
Since pure gapped ground states are expected to have modular strong spontaneously broken 1-form symmetries, 
ref.~\cite{sohal2024noisy} called density matrices with spontaneously broken strong symmetries that are not modular ``intrinsically mixed" topological orders. 
From our symmetry perspective, these intrinsically mixed topological orders demonstrate ``strong-to-weak" spontaneous symmetry breaking (SW-SSB) of 1-form symmetries \cite{xu2024average}.  

SW-SSB of 0-form symmetries has been studied in several recent papers \cite{lee2, ma2023topological, lessa2024strongtoweak, sala2024spontaneous, xu2024average, huang2024hydro, gu2024spontaneous, moharramipour2024symmetry, su2024emergent}. One reason why SW-SSB of higher form symmetries has not been as deeply explored is that under a commonly used equivalence relation on density matrices given by two-way finite-depth channel connectivity (see for example \cite{coser2019classification,sang2023mixed}, which we explain further in Sec.~\ref{smarkov}), density matrices that demonstrate higher form SW-SSB are often equivalent to ones that do not demonstrate SW-SSB, and therefore do not describe distinct phases. \footnote{More precisely, every ``intrinsically mixed" topological order demonstrates SW-SSB of bosonic or fermionic 1-form symmetries. A mixed state with strong symmetries of the form $\mathcal{C}\boxtimes\mathcal{T}$ where $\mathcal{C}$ is modular and $\mathcal{T}$ consists only of transparent bosons (which braid trivially with all other strong 1-form symmetries) is two-way channel connected to a state with modular strong symmetries $\mathcal{C}$, that is not intrinsically mixed\cite{ellison2024classification} A mixed state with SW-SSB of fermionic 1-form symmetries are also trivialized if we add fermionic ancillas.} This is the case in the decohered toric code example studied in several of the above references: the classical memory phase, which may demonstrate SW-SSB of 1-form symmetry,\footnote{At $p>p_c^{(2)}$ the density matrix demonstrates SW-SSB according to Sec.~\ref{ssymmetries}. $p_c^{(2)}>p_c$ of the classical memory (see Sec.~\ref{sdiscussion}).}. is two-way finite-depth channel connected to the infinite temperature state, if we are allowed to explicitly break the exact 1-form symmetry (see Appendix \ref{app:channel} for the explicit channel). Since we should allow explicit symmetry breaking for higher form symmetries (for pure states with parent Hamiltonians, we can explicitly break the 1-form symmetries while maintaining the topological order as long as we maintain the gap of the Hamiltonian), this means that the classical memory ``phase," and the SW-SSB phase, is actually trivial under the standard equivalence relation of two-way finite-depth channel connectivity. 

In this work, we use a slightly modified equivalence relation similar to the one proposed in Ref.~\cite{Sang2024TO}, based on finite R\'enyi-2 Markov length, that allows us to distinguish the SW-SSB phase and infinite temperature state as belonging in different phases, even with explicit symmetry breaking. Specifically, we say that $\rho_A\sim \rho_B$ if and only if there exists a finite time Lindbladian evolution between the two states with analytically varying, finite R\'enyi-2 Markov length. Ref. \cite{sang2024Markov} studied von Neumann Markov length, and showed that finite Lindbladian evolution preserving finite (von Neumann) Markov length implies two-way local channel connectivity. However, the converse does not necessarily hold, and is what allows us to obtain a finer classification. Because we use R\'{e}nyi-2 Markov length, we obtain in addition to the two-way local channel connectivity phase boundaries new phase boundaries that are not detected by von Neumann Markov length or other von Neumann quantum information measures. We discuss the comparison to von Neumann phase boundaries in Sec.~\ref{sdiscussion}. The ``continuously varying" requirement is to rule out situations where the Renyi-2 Markov length jumps discontinuously from one finite value to another finite value (we describe this possibility in more detail in Sec.~\ref{sdiscussion}). 

Unlike Ref.~\cite{Sang2024TO}, which uses von Neumann Markov length, which has nicer information-theoretic meaning, we choose to use R\'enyi-2 Markov length. This is because it is easier to compute and, as we will show in section \ref{smarkov}, its divergence appears to coincides nicely with correlation length divergence in the Choi state of the density matrix.\footnote{We expect that using the usual von Neumann Markov length can similarly distinguish the SW-SSB phase, and in fact the entire classical memory phase, from the infinite temperature state, with slightly different phase boundaries.  Ref.~\cite{Sang2024TO} will discuss the von Neumann version, using more information-theoretic methods.} Therefore, this classification naturally matches with a classification using the pure Choi states. More generally, it allows us to distinguish all intrinsically mixed topological orders from those with modular strong 1-form symmetries, which are equivalent to pure state topological orders. 

Equipped with the above definitions, we consider a natural setting to study mixed states that have been relatively unexplored: disordered systems. Specifically, given an ensemble of Hamiltonians with different disorder realizations, a density matrix can be constructed with the corresponding ensemble of ground states. We will investigate in detail the example of toric code with random vertex terms in section \ref{sex1}, which exhibits a robust SW-SSB/intrinsically mixed phase. In section \ref{sex2} we will discuss the more involved case of disordered toric code with random fields. Upon tuning the disorder strength, the system can host robust phases of matter demonstrating 1-form symmetry that are ``strong-to-trivial"-SSB (ST-SSB), ``strong-to-weak"-SSB (SW-SSB) and weakly symmetric (WS). In particular, we show that the density matrix describing the toric code with random field disorder can be mapped exactly to the decohered toric code at perturbative disorder amplitude.

Finally, we discuss subtleties and future directions in Sec.~\ref{sdiscussion}. We include the details of various exact and perturbative calculations in the Appendices; the techniques described there may be of independent interest.

\section{Preliminaries}\label{sprelim}

In Sec.~\ref{sdensity}, we will first provide a brief review on density matrices and quantum channels, and an equivalence relation on density matrices known as two-way channel connectivity. We will then discuss more precisely our slightly refined equivalence relation for density matrices in Sec.~\ref{smarkov}, and show that under our definition of equivalence (based on finite Lindbladian evolution preserving finite R\'enyi-2 Markov length), phases of density matrices map onto equivalence classes of Choi states.\footnote{As we will explain further in Sec.~\ref{smarkov}, the notion of equivalence class used here for pure states may not match in subtle ways with the notion of phase typically used for pure states.} The Choi state of a (possibly mixed) density matrix is a pure state in a doubled Hilbert space.\footnote{Sometimes the Choi state is defined with a normalization equal to the purity of the mixed state density matrix. In this work we will always refer to the normalized Choi state, which has norm 1. The normalization does not affect scaling of correlation functions.} Therefore, to classify phases of density matrices under our equivalence relation, we just need to classify Choi states. To classify topological phases of mixed states, we define in \ref{ssymmetries} strong and weak symmetries of density matrices and the meaning of these symmetries in their Choi states. 

An important difference between higher-form symmetries and 0-form symmetries is that higher-form symmetries can still emerge even when explicitly broken. Therefore, the corresponding topological phases are robust to generic perturbations. We will comment on emergent 1-form symmetries of the Choi state in Sec.~ \ref{semergent}, which correspond to, defined appropriately, emergent strong and weak symmetries of $\rho$.

\subsection{Density matrices and finite depth quantum channels}\label{sdensity}

While the definition of ``phase" in isolated quantum systems is well established, the definition of ``phase" in open quantum systems has only been considered in detail more recently. In order to define a phase for open quantum systems, one has to define an equivalence relation between density matrices. For pure states, the standard equivalence relation is that two states $|\psi\rangle$ and $|\psi'\rangle$ belong in the same phase if and only if $|\psi'\rangle=U|\psi\rangle$ where $U$ describes finite time evolution with a local Hamiltonian. One proposal for the analogous equivalence relation on mixed states, adopted by much of the recent literature \cite{de_Groot_2022, MaWangASPT, ma2023topological, sang2023mixed, sang2024Markov}, is as follows: two density matrices $\rho$ and $\rho'$ are said to belong in the same phase ($\rho\sim \rho'$) if and only if there exists a finite depth local quantum channel $\E[\cdot]$ connecting $\rho$ to $\rho'$ and another finite-depth local quantum channel $\E'[\cdot]$ connecting $\rho'$ to $\rho$:
\begin{equation}
    \rho\sim \rho':\exists \E[\cdot], \E'[\cdot] \qquad s.t \quad \E[\rho]=\rho',\quad \E'[\rho']=\rho
\end{equation}
If there exists such quantum channels $\E[\cdot]$ and $\E'[\cdot]$, then $\rho$ and $\rho'$ are said to be ``two-way finite-depth channel connected."
\begin{align}
\begin{split}
\E[\rho]&=\mathrm{Tr}_a(U^\dagger\rho\otimes |a\rangle\langle a| U)=\bigotimes_i\E_i[\rho]\\
\E_i[\rho]&=\sum_nK_{i,n}\rho K_{i,n}^\dagger
\end{split}
\end{align}
where $\{K_{i,n}\}$ are local Kraus operators satisfying $\sum_nK_{i,n}^\dagger K_{i,n}=1$. The continuous time version of a finite depth quantum channel is finite time Lindbladian evolution:
\begin{equation}
    \frac{d}{dt}\rho=\mathcal{L}\rho
\end{equation}
where $\mathcal{L}$ is the Lindblad superoperator (``Lindbladian").\footnote{Many examples of finite depth quantum channels require Lindbladian evolution with time that scales logarithmically with the system size. In this work, as in the literature, logarithmic time is regarded as finite time.} When $U$ does not couple the system with the ancillas, the Lindbladian evolution reduces to the usual Hamiltonian evolution: $\mathcal{L}\rho=-i[H,\rho]$. For incoherent local noise, the time $t$ is related to the decoherence strength. We give some examples of finite depth quantum channels and Lindbladian evolution in Appendix~\ref{sec:decohered}. 

A simple example of two-way finite-depth channel connectivity is the toric code model under an incoherent $Z$ noise channel, which is given by
\begin{equation}\label{dec}
    \rho\to(1-p)\rho+pZ_e\rho Z_e
\end{equation}
where $p\in[0,\frac{1}{2}]$ and $Z_e$ is a Pauli $Z$ operator on a single edge $e$. The action of this channel on every edge $e$ on the toric code ground state has been shown to be (approximately) reversible for sufficiently small $p$ , via the Petz recovery map \cite{Junge_2018, Kwon_2022, sang2024Markov}. Therefore, the corrupted toric code is in the same phase as the original toric code for sufficient small $p$. The physical intuition for the recoverability of the state at small $p$ is that for sufficiently small $p$, the amplitude of states with large Pauli $Z$ strings are very suppressed. Therefore, the logical state of the toric code on a torus can still be recovered. We will discuss the decohered toric code example in more details in the discussion section and appendix \ref{sec:decohered}. 

The classification of phases using two-way channel connectivity as the equivalence relation reduces to the usual classification of phases for pure states, and it has a nice operational meaning in terms of recoverability. However, it is too coarse for some purposes. For example, consider the following 2D state:
\begin{equation}\label{bp}
    \rho\propto\prod_p\left(\frac{1+B_p}{2}\right),
\end{equation}
where $B_p=\prod_{e\in p}Z_e$ are the standard plaquette operator in the toric code model. This is the state obtained from toric code by setting $p=\frac{1}{2}$ in (\ref{dec}). This state is known to have classical memory \cite{Chamon2007, Hamma_2009, TC_field} because it can be written as a sum of unentangled product states, but there are classical correlations between the states. As we will show below, the above state demonstrates SW-SSB of a 1-form symmetry generated by Pauli $Z$ loops, and has nontrivial  ``topological conditional mutual information" (we define this quantity in Sec.~\ref{sentanglementmain}). This state is also a simple example of an intrinsically mixed topological order (see Ref.~\cite{sohal2024noisy} for an in-depth discussion). However, this state is two-way channel connected to the infinite temperature state, as we show in Appendix~\ref{app:channel}, so under this equivalence relation it is trivial.

\subsection{Equivalence relation via analytic finite R\'enyi-2 Markov length}\label{smarkov}

In this work, we will use a slightly refined equivalence relation than two-way channel connectivity. To be precise, we will focus on topological phases, and will restrict to states that are Rényi-1 and R\'enyi-2 locally correlated, in the terminology of Ref.~\cite{ellison2024classification}. According to Ref.~\cite{ellison2024classification}, Rényi-1 locally correlated states satisfy
\begin{equation}\label{r1loc}
    \mathrm{Tr}(O_iO_j\rho)-\mathrm{Tr}(O_i\rho)\mathrm{Tr}(O_j\rho)\leq\mathcal{O}(|i-j|^{-\infty})
\end{equation}
where $O_i$ is an operator fully supported on site $i$ and $O_j$ is an operator fully supported site $j$. The above equation means that connected correlation functions decay superpolynomially in distance. R\'enyi-2 locally correlated satisfy a similar condition:
\begin{align}
\begin{split}\label{r2loc}
    &\frac{\mathrm{Tr}(\rho O_iO_j\rho O_j^\dagger O_i^\dagger)}{\mathrm{Tr}(\rho^2)}-\frac{\mathrm{Tr}(O_i\rho O_i^\dagger \rho)}{\mathrm{Tr}(\rho^2)}\frac{\mathrm{Tr}(O_j\rho O_j^\dagger\rho)}{\mathrm{Tr}(\rho^2)}\\
    &\leq\mathcal{O}(|i-j|^{-\infty})
\end{split}
\end{align}
The above two conditions are used to rule out GHZ-like states, related to spontaneous symmetry breaking of 0-form symmetries. We will see later that the above quantities correspond to connected correlation functions in the Choi state, so the above conditions ensure that connected correlation functions in the Choi state decay quickly. Topological phases can display nontrivial connected correlation functions on manifolds with nontrivial genus; in this section we restrict the discussion to the plane.\footnote{Note that analogous restrictions need to be used to define equivalence classes of pure states. For example, in order to say that the $\mathbb{Z}_2$ SSB phase is not a stable phase if we allow explicit breaking of $\mathbb{Z}_2$ symmetry, we must use the short-range entangled (fully polarized) basis for the ground state space. The GHZ state is not adiabatically connected to the product state even in the absence of $\mathbb{Z}_2$ symmetry.}

In this context, we say that two Rényi-1 and R\'enyi-2 locally correlated states $\rho$ and $\rho'$ belong in the same phase if and only if there exists paths, given by finite-time local Lindbladian evolutions, that connect $\rho\to\rho'$ and $\rho'\to\rho$ such that the R\'enyi-2 Markov length (which we will define below) remains finite along these paths. We will show that this definition of mixed state phases gives a classification that matches that given by equivalence classes of the Choi states, which we will define below. In particular, we will show that the R\'enyi-2 Markov length of a density matrix diverges if and only if there is a transition in the corresponding Choi state, marked by diverging correlation length. The analyticity requirement is to eliminate first-order transitions in the Choi state.
\footnote{While there may be a first order transition in Choi state, as we discuss further in Sec.~\ref{sdiscussion}, this may not be possible for Choi states corresponding to density matrices under local Lindbladian evolution.} While the discussion in this section is mostly on general grounds, we show in Appendix~\ref{si2tc} explicitly that the R\'enyi-2 Markov length diverges twice for toric code with $Z$ and $X$ noise, reflecting the fact that under this definition of equivalence, there are three phases: ST-SSB, SW-SSB, and WS.

Ref.~\cite{sang2024Markov} showed that if the (von Neumann) Markov length remains finite under the finite-time local Lindbladian evolution between two states, then we can always construct a reverse channel through the Petz recovery map \cite{Petz_map} to achieve a two-way finite-depth channel connection between the two states.\footnote{Some finite depth channels require time logarithmic in the system size when we formulate them using Lindbladian evolution. In alignment with the literature, we consider for log time ``finite".} However, the existence of two-way finite depth channels between two states does not imply that there exists finite-time local Lindbladian evolution preserving finite Markov length connecting the two states, as will be discussed in forthcoming work\cite{Sang2024TO} on von Neumann Markov length. According to our R\'enyi-2 results, we get a finer classification if we choose to use the above equivalence relation rather than two-way channel connectivity. For example, the state (\ref{bp}) is trivial under two-way channel connectivity because it is two-way channel connected to the infinite temperature state, but it is nontrivial under the equivalence relation of finite Lindbladian evolution preserving finite R\'enyi-2 Markov length. Note that using R\'enyi-2 quantities also slightly changes the phase boundaries. For example, for the toric code with incoherent noise, the R\'enyi-2 Markov length diverges when the Choi state undergoes a phase transition. The statistical model that describes the Choi state transition is the 2D classical Ising model rather than the 2D random bond Ising model (see Sec.~\ref{sdiscussion} and Fig.~\ref{fig:decohered} for a more in-depth discussion). Hence, the Choi state transition occurs at a slightly higher decoherence rate than that of the canonical purification, which detects two-way channel connectivity.

\begin{figure}[tbp]
\centering
\includegraphics[width=.5\columnwidth]{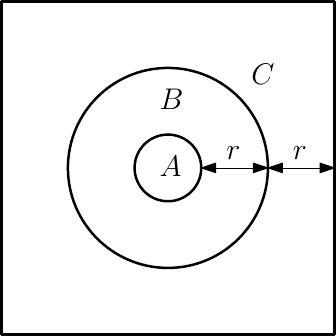} 
\caption{ The three regions on a plane used to defined the CMI. The radius of $A$ is an unimportant constant. We will take the width of $B$ and $C$, labeled by $r$, to infinity.}
\label{fig:CMI}
\end{figure}

\subsubsection{Definition of R\'enyi-2 Markov length}\label{sdefmarkov}

The Markov length gives the length scale for exponential decay of the conditional mutual information (CMI) in a particular geometry. The CMI $I(A:C|B)$ describes entanglement involving three subsets of the lattice $A,B,$ and $C$. Von Neumann $I(A:C|B)$ is defined as
\begin{equation}\label{neumanncmi}
I(A:C|B)=[S(AB)-S(B)]-[S(ABC)-S(BC)]
\end{equation}
where $S(X)$ is the von Neumann entanglement entropy between $X$ and its complement. For the particular choice of $A,B,$ and $C$ shown in Fig.~\ref{fig:CMI}, $I(A:C|B)$ is expected to decay exponential in $r$ for non-critical states\cite{sang2024Markov}. Specifically, in Fig.~\ref{fig:CMI}, $A$ is a disk of finite radius, $B$ is an annulus around $A$ of width $r$, and $C$ is the rest of the plane, also of width $r$. We will take the radius of $A$ to be an unimportant finite value and will take $r$ to infinity. $I(A:C|B)$ describes the information about $A$ contained in $BC$, that is not in $B$. Outside of critical points and gapless phases, we expect that $I(A:C|B)$ goes to zero exponentially: $I(A:C|B)\sim \mathrm{exp}(-r/\xi)$. Here, $\xi$ is referred as the von Neumann Markov length\cite{sang2024Markov}. 

We will use a R\'enyi-2 version of the CMI. Generally this quantity does not have nice information-theoretic properties because R\'enyi-2 entanglement entropy does not obey subadditivity or strong subadditivity \cite{Berta_2015a, Berta_2015b}. However, as we will show, the R\'enyi-2 Markov length has a nice connection to connected correlation functions in the corresponding Choi state. The R\'enyi-2 Markov length describes the decay of the R\'enyi-2 CMI, which we define by replacing von Neumann entanglement entropy in (\ref{neumanncmi}) be R\'enyi-2 entanglement entropies:\footnote{There are other definitions of Rényi CMI that have nicer information-theoretic properties \cite{Berta_2015a, Berta_2015b}. We will use the naive definition below due to its direct connection to the Choi state.}
\begin{equation}\label{Rényi2cmi}
I_2(A:C|B)=[S_2(AB)-S_2(B)]-[S_2(ABC)-S_2(BC)]. 
\end{equation}
We can write this explicitly as
\begin{equation}
I_2(A:C|B)=-\log\frac{\mathrm{Tr}[(\mathrm{Tr}_C(\rho))^2]\mathrm{Tr}[(\mathrm{Tr}_A(\rho))^2]}{\mathrm{Tr}[(\mathrm{Tr}_{AC}(\rho))^2]\mathrm{Tr}[\rho^2]}.
\end{equation}

We will show that this quantity fails to decay exponentially (so $\xi_2$ diverges) when the Choi state encounters second order phase transitions, i.e. when correlation length diverges. This suggests that phases of density matrices map onto equivalence classes of Choi states, which are connected components separated by diverging correlation length.

\subsubsection{Relation to the Choi state}\label{sRényichoi}

To derive the above result, the first step is to write the partial trace over a region $C$ of an operator $O$  by integrating it over unitaries supported on $C$ with the Haar measure. This gives $\mathrm{Tr}_C(O)$ tensored with the identity operator on $C$, appropriately normalized:
\begin{equation}
O|_{\bar{C}}=\int dU_C U_CO U_C^\dagger = \frac{1}{D_C}\mathrm{Tr}_C(O)\otimes \mathbf{1}_C,
\end{equation}
where $D_C$ is the Hilbert space dimension of $C$. $O|_{\bar{C}}$ is fully supported in the complement of $C$ because it commutes with all unitaries supported in $C$. This comes from the fact that the integration is invariant under $U_C\to U_C\cdot U$ for any $U$ supported in $C$. Then we can compute $\mathrm{Tr}(\mathrm{Tr}_C(O)^2)=D_C\mathrm{Tr}(O|_{\bar{C}}^2)$. 

For qubit systems, we can perform the ``averaging over unitaries" more explicitly by summing over Pauli strings (this also generalizes to qudit systems; we simply need a complete basis for the on-site operator algebra):
\begin{equation}\label{sumpauli}
O|_{\bar{C}}=\frac{1}{D_C^2}\sum_{\sigma_n\in C}\sigma_n O\sigma_n^\dagger
\end{equation}
where $\sigma_n$ runs over all operators of the form $\prod_i X_i\prod_j Z_j$ that are fully supported in $C$. 
Since $O|_{\bar{C}}$ commutes with all Pauli strings fully supported in $C$, it must therefore also commute with any operator fully supported in $C$. Furthermore, it is easy to see that the left and right sides of (\ref{sumpauli}) have the same trace, so if $O$ initially has unit trace then $O|_{\bar{C}}$ also has unit trace. Now we can write
\begin{align}
\begin{split}\label{DCtr}
D_C\mathrm{Tr}(O|_{\bar{C}}^2)&=\frac{1}{D_C^3}\sum_{\sigma_n,\sigma_m}\mathrm{Tr}(\sigma_nO\sigma_n^\dagger\sigma_m O \sigma_m^\dagger)\\
&=\frac{1}{D_C}\sum_{\sigma_n}\mathrm{Tr}(\sigma_n O \sigma_n^\dagger O). 
\end{split}
\end{align} 
The last line can be expressed as a sum of expectation values in the unnormalized Choi state $|\tilde{\rho}\rrangle$.

Specifically, a density matrix $\rho=\sum_i p_i |\psi_i\rangle \langle \psi_i|$ in a Hilbert space $\mathcal{H}$ can be mapped to its Choi state in a doubled Hilbert space $\mathcal{H}_u\otimes \mathcal{H}_l$. Here we use $u$ and $l$ to refer to the ``upper" and ``lower" Hilbert spaces which correspond to the ket and bra respectively. The (normalized) Choi state is defined as

\begin{equation}
|\rho\rrangle = \frac{1}{\sqrt{\mathrm{Tr}(\rho^2)}}\sum_n p_n |\psi_{n,u}\rangle \otimes |\psi_{n,l}^*\rangle,
\end{equation}

The unnormalized Choi state $|\tilde{\rho}\rrangle$ is the same as $|\rho\rrangle$ except with norm equal to the purity of $\rho$. If $\rho$ is a pure state, then $|\rho\rrangle$ is a tensor product: $|\rho\rrangle=|\psi_u\rangle|\psi_l^*\rangle$. Therefore, a pure topologically ordered density matrix maps onto the topological order stacked with its time reversal partner. If $\rho$ is not pure, then there is generally some entanglement between spins in the upper and lower Hilbert spaces. 

Returning to (\ref{DCtr}), can use the Choi state to write
\begin{equation}\label{Rényi2choi}
\mathrm{Tr}(\mathrm{Tr}_C(\rho)^2)=\frac{1}{D_C}\sum_{\sigma_n\in C}\llangle\tilde{\rho}|\sigma_{n,u}\sigma_{n,l}^\dagger|\tilde{\rho}\rrangle, 
\end{equation}

Note that in deriving the above equation, we used the observation that because $\sigma_{n,l}$ is a product of Pauli $X$ and $Z$ operators, its transpose is the same as the conjugate transpose. Using the expression above, we can write the R\'enyi-2 CMI in terms of expectation values in the Choi state:
\begin{align}
\begin{split}\label{I2choi}
&I_2(A:C|B)\\
&=-\log\frac{\sum_{\sigma_n\in C,\sigma_m\in A}\llangle\rho|\sigma_{n,u}\sigma_{n,l}^\dagger|\rho\rrangle\llangle\rho|\sigma_{m,u}\sigma_{m,l}^\dagger|\rho\rrangle}{\sum_{\sigma_n\in C,\sigma_m\in A}\llangle\rho|\sigma_{n,u}\sigma_{n,l}^\dagger\sigma_{m,u}\sigma_{m,l}^\dagger|\rho\rrangle}
\end{split}
\end{align}
where we simplified the expression using the normalized Choi state $|\rho\rrangle=\frac{1}{\sqrt{\mathrm{Tr}(\rho^2)}}|\tilde{\rho}\rrangle$. (\ref{I2choi}) is the main result of this section. It connects the R\'enyi-2 Markov length to connected correlation functions in the Choi state.

Note that the R\'enyi-2 locally correlated condition (\ref{r2loc}) implies that expectation values factorize superpolynomially quickly as we take the distance between $A$ and $C$ to infinity. This means that the numerator in (\ref{I2choi}) approaches the denominator of (\ref{I2choi}) very quickly as $r\to\infty$. In this case, it is easy to see that $I_{2}(A:C|B)$ decays superpolynomially with $r$; generically we expect $I_{2}(A:C|B)$ to decay exponentially with a finite Markov length. More precisely, let us denote 
\begin{align}  
\begin{split}
\Delta_{n,m}&=\llangle\rho|\sigma_{n,u}\sigma_{n,l}^\dagger\sigma_{m,u}\sigma_{m,l}^\dagger|\rho\rrangle\\
&-\llangle\rho|\sigma_{n,u}\sigma_{n,l}^\dagger|\rho\rrangle\llangle\rho|\sigma_{m,u}\sigma_{m,l}^\dagger|\rho\rrangle
\end{split}
\end{align}

Then 
\begin{align}
\begin{split}\label{i2expand}
    &I_2(A:C|B)\\
    &=-\log\left(1-\frac{\sum_{m,n}\Delta_{n,m}}{\sum_{\sigma_n\in C,\sigma_m\in A}\llangle\rho|\sigma_{n,u}\sigma_{n,l}^\dagger|\rho\rrangle\llangle\rho|\sigma_{m,u}\sigma_{m,l}^\dagger|\rho\rrangle}\right)\\
    &\sim\frac{\sum_{m,n}\Delta_{n,m}}{\sum_{\sigma_n\in C,\sigma_m\in A}\llangle\rho|\sigma_{n,u}\sigma_{n,l}^\dagger|\rho\rrangle\llangle\rho|\sigma_{m,u}\sigma_{m,l}^\dagger|\rho\rrangle}
\end{split}
\end{align}
where we expanded to leading order in $\frac{\sum_{m,n}\Delta_{n,m}}{\sum_{\sigma_n\in C,\sigma_m\in A}\llangle\rho|\sigma_{n,u}\sigma_{n,l}^\dagger|\rho\rrangle\llangle\rho|\sigma_{m,u}\sigma_{m,l}^\dagger|\rho\rrangle}$. 

Suppose that $|\rho\rrangle$ displays exponential decay of connected correlations. Then $\Delta_{n,m}=0$ if $\sigma_{n,u}\sigma_{n,l}^\dagger =\mathbf{1}$ or $\sigma_{m,u}\sigma_{m,l}^\dagger =\mathbf{1}$ and $\Delta_{n,m}$ decays exponentially in $r$ otherwise. Therefore, $I_2(A:C|B)$ seems to also decays exponentially in $r$, and we obtain a finite R\'enyi-2 Markov length. 

For example, if $\rho=|\psi\rangle\langle\psi|$ is a pure state so $|\rho\rrangle=|\psi\rangle|\psi\rangle$, then assuming that $\rho$ can be written as a product of $|A\cup B\cup C|$ local projectors \footnote{Note that if $|\psi\rangle$ is related by a local unitary $U$ to a fixed point state $\prod_i P_i$ where each $P_i$ is a local projector, then $|\psi\rangle$ can also be written as a product of local projectors: $|\psi\rangle\langle\psi|=\prod_i U^\dagger P_i U$.}, the denominator of (\ref{i2expand}) is of order $2^{|A|+|C|}$. On the other hand, the numerator is bounded (very loosely) by $(2^{|A|+|C|}-2^{|A|}-2^{|C|})e^{-\mu r}$ for some constant $\mu$, so $I_2(A:C|B)$ decays exponentially with $r$.  More generally when $\rho$ is mixed, then the denominator will typically be of order $e^{\nu (|A|+|C|)}$ for some $\nu$ (for example, if $\rho$ is the completely mixed state, $e^{\nu}=4$) and the numerator is loosely upper bounded by $(e^{\nu(|A|+|C|)}-e^{\nu|A|}-e^{\nu|C|})e^{-\mu r}$ so $I_2(A:C|B)$ is upper bounded by a function exponentially decaying in $r$.   If instead $\Delta_{n,m}$ decays with a power law for $\mathcal{O}(e^{\nu(|A|+|C|)})$ operators, then $I_2(A:C|B)$ cannot be fit to an exponential decay. Attempting to fit $I_2(A:C|B)$ to exponential decay in $r$ would lead to a diverging length scale, which is the diverging R\'enyi-2 Markov length.

Therefore, we expect that the R\'enyi-2 Markov length diverges whenever the Choi state encounters phase transitions where operators of the form $O_{u}O_l^\dagger$ have power-law decay of correlations. Note that since the expression (\ref{I2choi}) only has operators that are diagonal between the upper and lower Hilbert spaces, it is not bounded by decay of correlations of operators solely supported in a single Hilbert space, i.e. those of the form $\llangle\rho|\sigma_{n,u}\sigma_{m,u}|\rho\rrangle$. However, due to the $\mathrm{SWAP}^*$ symmetry of the Choi state (explained further in Sec.~\ref{semergent}), the above scenario where there is only a diverging correlation length in $\mathcal{H}_u$ or $\mathcal{H}_l$ is ruled out.

In the example of decohered toric code \cite{fan2024diagnostics, bao2023mixedstate, lee2, Zou_2023, sang2024Markov} (see Appendix~\ref{sec:decohered} for a review) where we put the state $|tc\rangle\langle tc|$ through an incoherent $Z$ noise channel (\ref{dec}) and $X$ noise channel (same as (\ref{dec}) except with $Z_e$ replaced by $X_e$), the Choi state is initially described by two copies of toric code $|\rho\rrangle=|tc\rangle|tc\rangle$. Under decoherence, the Choi state generically demonstrates two transitions, because there is a transition from two copies of toric code to a single copy, and then from a single copy to the trivial state. In Appendix~\ref{si2tc}, we show by explicit calculation in this example that $I_2(A:C|B)$ is related to the free energy of a point defect in the 2D classical Ising model, and $\xi_2$ diverges at two points.

\subsection{Strong and weak 1-form symmetries and order/disorder parameters}\label{ssymmetries}

Now that we have defined our equivalence relation on density matrices and showed that it maps onto equivalence classes of Choi states (connected components separated by transitions marked by diverging correlation length), we can consider the classification of such phases. The classification of topological phases (without any additional 0-form symmetry) is given by spontaneous symmetry breaking patterns of 1-form symmetries. As we will show below, 1-form symmetries of the Choi state map onto ``strong" and ``weak" 1-form symmetries of the original mixed state $\rho$. 

In this work, we will only consider abelian 1-form symmetries in 2+1d in bosonic systems, but the definitions in this section generalize straightforwardly to abelian higher-form symmetries in other spacetime dimensions. For a pure state to have a symmetry $G$ with a unitary representation $\{U_g\}$, it must be an eigenstate of each symmetry operator: $U_g|\psi\rangle=e^{i\theta_g}|\psi\rangle$ for all $g\in G$. A density matrix respects a strong symmetry $G$ if \cite{de_Groot_2022}
\begin{equation}\label{strong}
    U_g\rho=e^{i\theta_g}\rho,\qquad\forall g\in G.
\end{equation}
By comparison, a density matrix respects a weak unitary symmetry $G$ if 
\begin{equation}\label{weak}
    U_g\rho U_g^\dagger =\rho,\qquad\forall g\in G.
\end{equation}
Every strong symmetry is also a weak symmetry, but the converse does not hold. When a density matrix is weakly symmetric, it can be diagonalized in the eigenbasis of $U_g$. We can write $\rho=\sum_{i}p_i|\psi_i\rangle\langle\psi_i|$ where $\{|\psi_i\rangle\}$ are eigenstates of $U_g$. In this basis, for the weak symmetry to also be a strong symmetry, the charge of each eigenstate with nonzero $p_i$ must be the same: $\langle\psi_i|U_g|\psi_i\rangle = \langle\psi_j|U_g|\psi_j\rangle$ for all $g\in G$. For pure states, strong and weak symmetries are the same. 

To motivate the definition for order/disorder parameters of strong/weak 1-form symmetries, let's warm up by considering a simple example of Ising symmetry on a chain $U=\prod_i X_i$. In the pure state case, the order parameter is the two-point correlator $\langle Z_i Z_j\rangle$ since $Z_i$ is charged under $U$ but the product $Z_iZ_j$ is neutral (commutes with $U$).  The disorder parameter is simply $U$ restricted to a subsystem: the open string $\langle \prod_{i\in A} X_i \rangle$ where $A$ is an interval on the chain. 
Now for the Choi state of the mixed state, the weak Ising symmetry acts as
\be
U_{u} U_{l}^T|\rho\rrangle=|\rho\rrangle
\ee
where $u, l$ label the upper and lower Hilbert spaces corresponding to the ket and bra, respectively. The order parameter can be either two point correlator in the upper Hilbert space $\propto \llangle \rho | Z_{u,i}Z_{u,j}|\rho\rrangle$ or in the lower Hilbert space $\propto \llangle \rho | Z_{l,i}Z_{l,j}|\rho\rrangle$, because neither $Z_{u,i}$ nor $Z_{u,j}$ commutes with $U_u U_l^T$, but both $Z_{u,i}Z_{u,j}$ and $Z_{l,i}Z_{l,j}$ commute with $U_u U_l^T$. Since only operators from one of the two Hilbert spaces is involved, we label the order parameter $O_1$ with subscript $1$. We cannot use $O_2$ as an order parameter because $Z_{u,i}Z_{l,i}$ is not charged under $U_u U_l^T$. By comparison, the disorder parameter is the symmetry action restricted to a subsystem (a $d$-ball in $d$ spatial dimensions). Since the weak symmetry acts on both Hilbert spaces, the disorder parameter lives on both Hilbert spaces as well, which is $\propto \llangle \rho|\prod_{i\in A} X_{u,i} X_{l,i}|\rho\rrangle$, which explains the subscript $2$ in $D_2.$ Following the similar logic, the disorder parameter for the strong symmetry is $D_1$. Unlike for the weak symmetry, for the strong symmetry $O_2\propto \llangle\rho| Z_{u,i}Z_{u,j} Z_{l,i}Z_{l,j} |\rho\rrangle$ is a valid order parameter. This is because $Z_{u,i}Z_{l,i}$ is charged under the strong symmetry (it fails to commute with $U_u$). $O_1$ also serves as an order parameter for the strong symmetry, so the strong symmetry is SSB if either $O_2$ or $O_1$ is nonzero (for a symmetry to be SSB, it suffices that \emph{any} order parameter is nonzero).

In this work, we consider 1-form symmetries in 2+1d, i.e., the operators $U_g$ are closed string operators. In a pure state topological order, the 1-form symmetries are spontaneously broken. Spontaneous breaking of a 1-form symmetry $U_g$ corresponding to an anyon $g$ that can be detected by an order parameter. This order parameter is given by the expectation value of a closed string operator for another anyon that braids nontrivially with $g$. 

Strong and weak symmetries translate to different 1-form symmetries of the Choi state $|\rho\rrangle$, and the strong/weak order and disorder parameters map onto usual pure state order and disorder parameters of the Choi state. 

The strong symmetry (\ref{strong}) acts on only one side of the density matrix, so in the doubled Hilbert space, it only acts on $\mathcal{H}_{u}$ or $\mathcal{H}_{l}$:  
\begin{equation}
U_{g,u} |\rho\rrangle  \propto |\rho\rrangle\qquad U_{g,l}^T |\rho\rrangle  \propto |\rho\rrangle.
\end{equation}

For 1-form symmetries, these correspond to closed anyon loops that live fully in $\mathcal{H}_u$ or $\mathcal{H}_l$. A density matrix with a weak symmetry (\ref{weak}) maps onto a state satisfying
\begin{equation}
U_{g,u} U_{g,l}^T|\rho\rrangle=|\rho\rrangle,\quad g\in G.
\end{equation}

A weak 1-form symmetry generator corresponds to a closed anyon loop living in the upper Hilbert space and the time reversed anyon loop in the lower Hilbert space. 
Thus we observe that the strong and weak symmetries behave like conventional symmetries in the doubled Hilbert space. It follows that we can define the following order parameter for the strong symmetry: 
\begin{equation}
O_2=\lim\limits_{r\to\infty}\llangle\rho| U_{h,u}U_{h,l}^T|\rho\rrangle.
\end{equation}
where $U_{h,u(l)}$ is a closed string operator in $\mathcal{H}_u$ ($\mathcal{H}_l$) for an anyon $h$ that braids nontrivially with $g$. $r$ denotes the radius of the loop on which $U_{h,u(l)}$ is supported. 
\begin{equation}   O_2=\lim\limits_{r\to\infty}\frac{\mathrm{Tr}(U_h\rho  U_h^\dagger\rho)}{\mathrm{Tr}(\rho^2)}.
\label{eq:def_O2}
\end{equation}
The disorder parameter for strong symmetry is obtained by restricting the closed string operator $U_g$ to a segment of it, 
\begin{equation}   
D_1=\lim\limits_{|i-j|\to\infty}\llangle\rho| U_{g,u}(i,j)|\rho\rrangle,
\end{equation}
where $U_{g,u}(i,j)$ is the symmetry operator $U_{g,u}$ restricted to an open string with endpoints at $i,j$. In the original Hilbert space, it is given by
\begin{equation}
D_1=\lim\limits_{|i-j|\to\infty}\frac{\mathrm{Tr}(\rho U_{g}(i,j)\rho)}{\mathrm{Tr}(\rho^2)}.
\label{eq:def_D1}
\end{equation} 

Using the Choi state, it is straightforward to define order and disorder parameters for the weak symmetry:
\begin{equation}
O_1=\lim\limits_{r\to\infty}\llangle\rho| U_{h,u}|\rho\rrangle
\label{eq:def_O1}
\end{equation}
\begin{equation}
    D_2=\lim\limits_{|i-j|\to\infty}\llangle\rho| U_{g,u}(i,j)U_{g,l}^T(i,j)|\rho\rrangle
\end{equation}
Note that $O_1$ also serves as an order parameter for the strong symmetry. However, if $O_1$ is not long-ranged, this does not preclude the strong symmetry from being spontaneously broken; there just needs to be \emph{some} long-ranged order parameter, and $O_2$ might be long-ranged. $O_1$ and $D_2$ can also be calculated in the single Hilbert space:
\begin{equation}
O_1=\lim\limits_{r\to\infty}\frac{\mathrm{Tr}(\rho U_{h}\rho)}{\mathrm{Tr}(\rho^2)},
\end{equation}
\begin{equation}\label{d2def}
    D_2=\lim\limits_{|i-j|\to\infty}\frac{\mathrm{Tr}(U_g(i,j) \rho U_g^\dagger(i,j)\rho)}{\mathrm{Tr}(\rho^2)}.
\end{equation}

\subsection{Emergent 1-form symmetries}\label{semergent}

It is well known that pure state topological order is robust to generic small perturbations, even if the perturbations do not preserve the 1-form symmetries exactly. A small perturbation of a pure state $|\psi\rangle$ takes $|\psi\rangle\to |\tilde{\psi}\rangle=\mathcal{U}|\psi\rangle$ where $\mathcal{U}$ describes finite time-evolution by a local Hamiltonian. The perturbed state does not in general satisfy the exact symmetry condition (\ref{strong}). However, for sufficiently small perturbations (for example when the parent Hamiltonian is modified in a way that preserves the gap), there are \emph{emergent} 1-form symmetries. For a pure state, an emergent 1-form symmetry is a dressed unitary string operator $\tilde{U}_g=\mathcal{U}U_g\mathcal{U}^\dagger$ that satisfies the same algebraic relations as the original one, that is superpolynomially localized along the support of $U_g$ \cite{Hastings2011, cherman2024emergent}. Clearly, $\tilde{U}_g$ satisfies
\begin{equation}
    \tilde{U}_g|\tilde{\psi}\rangle\propto|\tilde{\psi}\rangle.
\end{equation}

Note that for a pure state, even if $U_g$ commutes with the parent Hamiltonian of $|\psi\rangle$, $\tilde{U}_g$ does not in general commute with the parent Hamiltonian of $|\tilde{\psi}\rangle$. In this sense, it is ``emergent." The corresponding Choi state of a pure state has separate (emergent) 1-form symmetries, $\tilde{U}_{g,u}$ and $\tilde{U}_{g,l}$, acting on the upper and lower Hilbert spaces. 

For mixed states, in contrast, small perturbations include finite-depth quantum channels \cite{sang2023mixed}, which in the Choi state representation
generically lead to coupling between the two Hilbert spaces. $|\psi\rangle|\psi^*\rangle$ gets mapped to $ \mathcal{U}_{ul}|\psi\rangle|\psi^*\rangle$ where $\mathcal{U}_{ul}$ couples the upper and lower layers.
As a result, the dressed 1-form symmetry generators get smeared between the two Hilbert spaces, and are no longer supported in a single Hilbert space. 
Therefore, the strong symmetry condition (\ref{strong}) may not be satisfied for any unitary, superpolynomially localized 1-form symmetry generator. In other words, there is no dressed unitary $\tilde{U}_g$ satisfying $\tilde{U}_g\rho\propto\rho$. Therefore, emergent 1-form symmetries of the Choi state do not translate in the most naive way to emergent strong 1-form symmetries of the density matrix $\rho$.

To get a definition of emergent strong 1-form symmetries of $\rho$, we must include ancillas and assert our finite R\'enyi-2 Markov length condition. However, since we already showed in Sec.~\ref{smarkov} that using our equivalence relation phases of density matrices are in one-to-one correspondence with phases of Choi states, and topological phases of Choi states are classified by the algebraic properties of their emergent 1-form symmetries, it suffices to define emergent 1-form symmetries of Choi states. These are dressed unitary string operators that are superpolynomially localized along closed loops, with the same algebraic relations as the original string operators, that satisfy
\begin{equation}
\tilde{U}_g|\rho\rrangle\propto|\rho\rrangle
\end{equation}

We define the topological order of the Choi state by the algebraic relations (braiding, fusion, etc) of the emergent 1-form symmetries, for which $|\rho\rrangle$ is an exact eigenstate. The phase of the density matrix then follows from the phase of its Choi state.\footnote{We do not assume that the definition of phase for the Choi state matches in this context exactly with the definition for usual pure gapped ground states. This is because the action of finite time Lindbladian evolution preserving finite Markov length on a density matrix, when translated to a unitary action on the Choi state, might not be finite time evolution of a local Hamiltonian. Here, we simply say that two Choi states are in the same ``phase" if they can be connected by action corresponding to finite Lindbladian evolution, i.e. a sequence of infinitesimal locally generated non-unitary evolutions appropriately normalized, without a diverging correlation length.} 

Note that even if the two layers are coupled, the Choi state always has an antiunitary symmetry $\mathrm{SWAP}^*$ that swaps the states in $\mathcal{H}_u$ and $\mathcal{H}_l$ and performs complex conjugation. Using this symmetry, we can distinguish between strong and weak symmetries in the Choi state even when the strong symmetries are smeared across both layers. (Here we slightly abuse notation. The Choi state is a pure state, so it only has strong symmetries. However, in considering the Choi state as a representation of a density matrix, we can distinguish symmetries that correspond to strong and weak symmetries of the density matrix). We define emergent weak symmetries in the Choi representation as emergent 1-form symmetries that are invariant under $\mathrm{SWAP}^*$ and emergent strong symmetries as those that are permuted under $\mathrm{SWAP}^*$.

\subsubsection{Entanglement properties}\label{sentanglementmain}
\begin{figure}[htbp]
\centering
\includegraphics[width=.5\columnwidth]{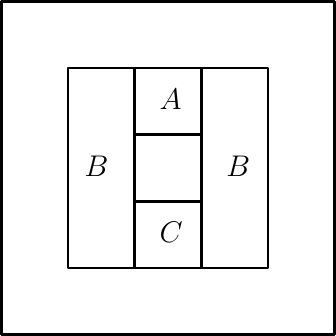} 
\caption{ The topological CMI is given by the combination of entanglement entropies in (\ref{neumanncmi}), with $A,B,C$ as illustrated above. For pure states, it just gives twice the TEE.}
\label{fig:lw}
\end{figure}

In pure states, spontaneous breaking of 1-form symmetries in pure states leads to long-range entanglement, which can be characterized by topological entanglement entropy (TEE) \cite{Kitaev_2006, Levin_2006}. Similarly, spontaneous breaking of strong and weak 1-form symmetries in mixed states can also be diagnosed using entanglement properties. Because our equivalence relation relates phases of density matrices to equivalence classes of Choi states, we can diagnose the phase using TEE of the Choi state, which is the CMI computed with $A,B,C$ in the Levin-Wen partitions illustrated in Fig.~\ref{fig:lw}. The physical intuition behind the TEE computed in the Levin-Wen partitions, for Abelian topological orders, is that it counts the number of string operators with expectation value 1 wrapping the hole surrounded by $A\cup B\cup C$, that are not generated by local operators with expectation value 1. It therefore counts the number of spontaneously broken 1-form symmetries, which may be emergent. 

We can also consider the R\'enyi-2 CMI of the original mixed density matrix computed with the Levin-Wen partitions \cite{Levin_2006}. We will call the R\'enyi-2 CMI of the original mixed state, computed in the Levin-Wen partitions, the LW CMI $\gamma(\rho)$ rather than TEE. This is because for mixed states, R\'enyi-2 CMI also includes classical correlations. At fixed points, it is easy to see that these two quantities match. For example, for a pure state, the LW CMI is simply twice the TEE of the pure state, and the TEE of the Choi state is also twice the TEE of the original pure state. We compute these quantities explicitly at some fixed points in Appendix~\ref{sentanglement}. We hypothesize that in the thermodynamic limit, these two quantities match even away from the fixed point. The TEE is independent of the Rényi index \cite{PhysRevLett.103.261601}, so it matches with the much more commonly studied von Neumann TEE. We therefore hypothesize that the LW CMI is also independent of Rényi index.

\section{Phases of mixed states: $\mathbb{Z}_2\times\mathbb{Z}_2$ 1-form symmetry}\label{sphases}

For concreteness we will now specialize to the case where $G=\mathbb{Z}_2\times\mathbb{Z}_2$, where the two $\mathbb{Z}_2$ 1-form symmetries have a mixed anomaly (specifically, a mutual braiding phase of -1). The strong symmetries describe the spontaneously broken symmetries of the well-known pure state toric code model. Including the weak symmetries, there are four symmetry generators. We will denote the two strong symmetry generators by the strong Wilson and 't Hooft symmetries. At the fixed point, when the symmetries are exact, we can write an explicit model where these correspond to closed loops of Pauli $Z$ and $X$ operators respectively. Similarly, we will denote the two weak symmetry generators by the weak Wilson and 't Hooft symmetries. 

The Choi state of a toric code ground state describes two copies of toric code. Therefore, there are three phases that can be obtained from the parent toric code state, corresponding to when the Choi state describes two copies of toric code (TO), a single copy of toric code (intrinsically mixed), or a trivial state. In the rest of this paper, we will study in depth various examples of disordered systems that give rise to density matrices in each of the above phases.

To give some intuition for these phases, we can examine their fixed points. A fixed point model for the toric code phase is simply the pure toric code state. A parent Hamiltonian is given by
\begin{equation}\label{tcham}
    H=-\sum_vA_v-\sum_pB_p
\end{equation}
where $A_v=\prod_{e\in v}X_e$ are vertex terms and $B_p=\prod_{e\in p}Z_e$ are plaquette terms, each of which are four-body on the square lattice. On a sphere, the unique ground state of the above Hamiltonian is 
\begin{equation}\label{puretc}
    \rho=\prod_p\left(\frac{1+B_p}{2}\right)\prod_v\left(\frac{1+A_v}{2}\right)
\end{equation}
The factor of four comes from the Euler characteristic of a discretized sphere. It is easy to see that $\rho$ has the following order and disorder parameters:
\begin{equation}
    O_{1,x}=1\qquad D_{1,z}=D_{2,z}=0
\end{equation}
where the extra subscript $z,x$ refers to the Wilson and 't Hooft symmetries respectively. All of the strong and weak symmetries are spontaneously broken. The density matrix has LW CMI $2\log2$ and the Choi state has TEE $2\log 2$. 

The intrinsically mixed-state fixed point is described by
\begin{equation}\label{SW-SSBz2}
    \rho=\frac{1}{2^{N-1}}\prod_p\left(\frac{1+B_p}{2}\right),
\end{equation}
where $N$ is the number of vertices. 
Computing the order and disorder parameters in this state gives
\begin{align}
\begin{split}
    &O_{2,x}=1\qquad D_{1,z}=0\\
    &O_{1,x}=0\qquad D_{2,z}=1\\
\end{split}
\end{align}
The density matrix has a strong Wilson symmetry which is spontaneously broken down to a weak symmetry. Meanwhile, the density matrix only respects a weak 't Hooft symmetry, which is also spontaneously broken. The density matrix has LW CMI $\log2$ and the Choi state has TEE $\log 2$. Because the strong $\mathbb{Z}_2$ 1-form symmetry is not modular (there is no other strong 1-form symmetries that braid non-trivially with it), this density matrix describes an intrinsically mixed state.

We note that there is a close relationship between SW-SSB of 0-form symmetry and that of 1-form symmetry. In (2+1)$d$, gauging a 0-form discrete finite global symmetry produces a dual 1-form symmetry generated by the Wilson loop operators of the original symmetry. 
The above state can be obtained from a density matrix demonstrating SW-SSB of a 0-form non-anomalous $\mathbb{Z}_2$ symmetry such as $\rho\propto1+\prod_vX_v$ where $X_v$ is a Pauli $X$ operator acting on a vertex $v$ (this is simply the 2+1$d$ version of the 1+1$d$ example in Ref.~\cite{ma2023topological}). Eq. (\ref{SW-SSBz2}) can be obtained by gauging the $\mathbb{Z}_2$ symmetry $\prod_vX_v$. This method for constructing 1-form SW-SSB states starting from 0-form SW-SSB states of non-anomalous 0-form symmetry can be generalized to other symmetry groups. See Sec.~\ref{sising} for more discussion on the relationship between the topological phases discussed here and Ising spin glass literature.

Finally, the trivial fixed point is described by
\begin{equation}\label{inftemp}
\rho=\frac{1}{2^{2N-2}}\mathbf{1}
\end{equation}

It only has weak Wilson and 't Hooft symmetries, and is symmetric under both of them. The density matrix has LW CMI $0$ and the Choi state has TEE $0$. Because the Choi state for the intrinsically mixed phase describes a single copy of toric code and the Choi state for the trivial phase describes a trivial topological order, then there must be a phase transition marked by diverging correlation length in any interpolation between the two states. It follows that the two density matrices are not equivalent via the definition in Sec.~\ref{smarkov}; the intrinsically mixed phase is a distinct phase from the trivial phase under this equivalence relation.  

Next, we will consider ensembles of disordered toric code ground states. Each one of the mixed states we obtain, which depend on the type and amplitude of disorder, will be equivalent to one of the above fixed points on the sphere. In particular, we will obtain examples of intrinsically mixed density matrices from disordered ensembles.

\section{Example: random vertex term}\label{sex1}

We now turn to our first example of a mixed state from an ensemble of disordered Hamiltonians. We consider an ensemble of Hamiltonians $\{H(\{\lambda_{i}\})\}$ where each Hamiltonian is described by a different set of random couplings $\{\lambda_{i}\}$. This ensemble of Hamiltonians comes with an ensemble of ground states. For simplicity, we will consider the state on a sphere, so every Hamiltonian has a unique ground state (disregarding accidental degeneracies). We then build a density matrix by putting these ground states in an incoherent superposition. 

For the particular case at hand, every disorder realization is given by a toric code Hamiltonian (\ref{tcham}) with a random coefficient in front of the vertex term:
\begin{equation}
    H(\{\lambda_{A,v}\})=-\sum \lambda_{B,p}B_p-\sum\lambda_{A,v}A_v
\end{equation}
We take the coefficients $\{\lambda_{A,v}\}$ from a Gaussian distribution with mean 1 and variance $\Delta_A^2$. For now, we will take $\lambda_{B,p}=\lambda_B=1$ for all plaquettes; more general $\{\lambda_{B,p}\}$ will be considered in Sec. \ref{srandplaquette}. Note that the Hamiltonian $H(\{\lambda_{A,v}\})$ for every disorder realization is exactly solvable, because the terms all mutually commute. Furthermore, the ground state for every disorder realization lies in the toric code phase. On the sphere, the unique ground state is given explicitly by the projector
\begin{align}
\begin{split}
    &|\psi_0(\{\lambda_{A,v}\})\rangle\langle\psi_0(\{\lambda_{A,v}\})|\\
    &=
    \prod_p\left(\frac{1+B_p}{2}\right)\prod_{v\neq v_0}\left(\frac{1+\mathrm{sign}(\lambda_{A,v})A_v}{2}\right),
\end{split}
\end{align}
where $v_0$ is a fixed vertex where the corresponding value of $A_{v_0}$ is determined from the identity $A_{v_0}=\prod_{v\neq v_0}A_v$. 
The ground state is identical to the ideal toric code ground state, except there is a probability that depends on $\Delta_A^2$ for a given vertex $v$ to have $\lambda_{A,v}<0$. The ground state then satisfies $A_v|\psi_0(\{\lambda_{A,v}\})\rangle=-|\psi_0(\{\lambda_{A,v}\})\rangle$ for that vertex. 

In this section, we will first show that the density matrix constructed by summing over all disorder realizations takes the form 
\begin{equation}\label{rhoAv}
    \rho(\Delta_A^2)=\frac{1}{[2\cosh(\beta_A)]^{N-1}}\prod_p\left(\frac{1+B_p}{2}\right)\prod_{v\neq v_0} e^{\beta_A A_v}
\end{equation}
where the effective inverse temperature $\beta_A$ is determined by $\Delta_A^2$, and $N$ is the number of vertices. Note that this result depends not only on the amplitude but also the distribution of the disorder; we will show in Sec.~\ref{srhoAv} that if we instead use bimodal disorder, we only obtain density matrices of the form (\ref{rhoAv}) with $\beta=0$ or $\beta=\infty$.

Before we proceed further, it is worth noting that the same density matrix can also be found in (1) the toric code at finite temperature with the coefficient $\lambda_B$ of the $B_p$ term taken to infinity and (2) toric code systems prepared by finite depth circuit, measurement, and error correction (see Refs. \cite{Chamon2007, Hamma_2009, iqbal2024topological, Tantivasadakarn_2024, lee2022decoding, Zhu_2023, Iqbal_2023, Iqbal_2024}), if there is a nonzero rate of measurement error. In the latter setting, $B_p=1$ automatically for all plaquettes due to the choice of measurement basis, as long as the finite-depth circuit has no error. However, $A_v=\pm 1$ depending on the outcome of the measurement on the ancillas. One must then correct for the $A_v=-1$ vertices by appropriate application of string operators. However, if there is a measurement error where the sign of an ancilla measurement is read off incorrectly, then one generally obtains a toric code state with a small percentage of vertices with $A_v=-1$. It follows from the results in this section that for any nonzero rate of measurement error, the resulting density matrix only forms a classical memory, and demonstrates SW-SSB. This gives some physical intuition for why measurement error, even at small probability, is much more detrimental than incoherent noise.

In the following sections, we will show that the density matrix in (\ref{rhoAv}) demonstrates a SW-SSB phase for any nonzero value of $\Delta_A^2$. In this phase, the density matrix is equivalent to (\ref{SW-SSBz2}). Furthermore, it always has CMI as defined in section \ref{sentanglementmain} of $\log 2$, and the corresponding Choi state has TEE $\log 2$. In addition, we will derive an exact path of gapped, local Hamiltonians interpolating between a parent Hamiltonian of the Choi state of (\ref{rhoAv}) and that of a pure toric code state with a decoupled product state, to show explicitly that these two Choi states are equivalent.

\subsection{Mixed state density matrix and CMI}\label{srhoAv}

In this section, we will show that the density matrix given by the normalized sum of ground state projectors with different disorder realizations $\{\lambda_{A,v}\}$ (which are orthogonal in the thermodynamic limit) is given by (\ref{rhoAv}). We will also compute the R\'enyi-2 conditional mutual information for (\ref{rhoAv}) explicitly. We note that the von Neumann CMI was computed in Ref. \cite{sang2024Markov}, but we will show that the R\'enyi-2 CMI calculation is simple and produces the same result.

We begin by deriving the density matrix. Upon averaging over disorder realizations, for any given vertex $v\neq v_0$ there is a probability $\frac{1}{2}\left[1\pm \mathrm{erf}\left(\frac{1}{\sqrt{2}\Delta_A}\right)\right]$ to get $\mathrm{sign}(\lambda_{A,v})=\pm 1$. Here, the error function $\mathrm{erf}(x)$ is defined as the integral over the Gaussian distribution from $0$ to $x$, normalized so that $\mathrm{lim}_{x\to\infty}\mathrm{erf}(x)=1$. For any given vertex, this probability distribution gives the following density matrix 
\begin{align}
\begin{split}
    \rho_v=\sum_{s=\pm 1} & P [\mathrm{sign}(\lambda_{A,v})=s]\left(\frac{1+s A_v}{2}\right)\\
    =\ & \frac{1+\mathrm{erf}\left(\frac{1}{\sqrt{2}\Delta_A}\right)A_v}{2}.
\end{split}
\end{align}
Therefore, summing over disorder realizations gives (\ref{rhoAv}) with
\begin{equation}
    \mathrm{tanh}(\beta_A)=\mathrm{erf}\left(\frac{1}{\sqrt{2}\Delta_A}\right),
\end{equation}
where have used the identity $e^{\beta_A O}=\cosh(\beta_A)+\sinh(\beta_A) O$ for any operator $O$ that squares to the identity. 
Note that as $\Delta_A\to0,$ $\beta_A\to \infty$ returns to the usual toric code model, while as $\Delta_A\to\infty,\beta_A\to 0$ as expected.

Interestingly, there is a sharp difference if we use a bimodal distribution instead of a Gaussian distribution. If instead we use a bimodal distribution $\lambda_{A,v}=1+\delta\lambda_{A,v}$ with $\delta\lambda_{A,v}=\pm \lambda$, then $\rho(\lambda)$ is the pure toric code ground state (\ref{puretc}) for all $\lambda<1$ and the $\beta_A\to0$ state (\ref{SW-SSBz2}) for all $\lambda>1$ (we do not consider the case $\lambda=1$ because then every Hamiltonian in the ensemble has an extensive degeneracy). Only with Gaussian disorder can we realize density matrices with $0<\beta_A<\infty$.

We will now compute the LW CMI of the density matrix (\ref{rhoAv}), which can be carried out exactly because (\ref{rhoAv}) is a product of commuting terms. We define the R\'enyi-2 LW CMI using the partitions in Fig.~\ref{fig:lw}. To compute the R\'enyi-2 entanglement entropy between a region and its complement, it is helpful to expand the density matrix in terms of Pauli strings. We have 
\begin{equation}
    \rho(\beta_A)=\frac{1}{2^{2N-1}}\prod_p(1+B_p)\prod_{v\neq v_0}\left(1+\tanh(\beta_A)A_v\right). 
\end{equation}
If we expand out the product above and trace out a region $\bar{R}$ which is the complement of a simply connected region $R$, then any nontrivial term with support in $\bar{R}$ would be removed by the trace because the trace of every Pauli operator is zero. Therefore, for any simply connected region $R$ which, without loss of generality, we assume to not contain $v_0$,
\begin{align}
\begin{split}
\rho_R(\beta_A)&=\mathrm{Tr}_{\bar{R}}\left[\rho(\beta_A)\right]\\
&=\frac{1}{2^{2N_e}}\prod_{p\in R}\left(1+B_p\right)\prod_{v\in R}(1+\tanh(\beta_A)A_v), 
\end{split}
\end{align}
where the product is over plaquettes and vertices fully supported in $R$ and {$N_e$} is the number of edges in $R$. It follows that
\begin{equation}\label{entropydisk}
    \mathrm{Tr}\left[\rho_R(\beta_A)^2\right]=\frac{2^{N_p}}{2^{N_e}}\left(1+\tanh^2(\beta_A)\right)^{N_v}.
\end{equation}
where $N_p (N_v)$ is the number of plaquettes (vertices) fully supported in $R$.

If instead we choose $R$ to be a non-simply connected region $R=A\cup B\cup C$ as defined in Fig.~\ref{fig:lw}, there are additional operators that can survive the trace: In the expansion of $\rho_R(\beta_A)$, not only are there closed contractible loops of Pauli $X$ and $Z$ operators obtained from products of $B_p$ and $A_v$ operators fully supported in $R$, but there are also loop operators that wind around the hole in $R$. This means that
\begin{align}
\begin{split}\label{rhohole}
    &\rho_R(\beta_A)=\frac{1}{2^{N_e}}\prod_{p\in R}\left(1+B_p\right)\left(1+\prod_{e\in\partial h} Z_e\right)\\
    &\times\prod_{v\in R}(1+\tanh(\beta_A)A_v)\left(1+\tanh^{|h|}(\beta_A)\prod_{e\in\partial h} X_e\right)
\end{split}
\end{align}
where $|h|$ is the area of the hole in the interior of the region $R$ (i.e. the square enclosed by $A\cup B\cup C$ in Fig.~\ref{fig:lw}). The $\prod_{e\in\partial h} X_e$ operator comes from the product of all of the $A_v$ terms inside the hole, which produces an operator fully supported in $R$. Similarly, the $\prod_{e\in\partial h} Z_e$ operator comes from the product of all the $B_p$ terms inside the hole. 

From (\ref{rhohole}), we can easily compute, 
\begin{align}
\begin{split}\label{entropyannulus}
    & \mathrm{Tr}[\rho_R(\beta_A)^2]\\
    =\ & \frac{2^{N_p+1}}{2^{N_e}}\left(1+\tanh^2(\beta_A)\right)^{N_v}
    \left(1+\tanh^{2|h|}(\beta_A)\right).
\end{split}
\end{align}

Using (\ref{entropydisk}) and (\ref{entropyannulus}), we can compute the CMI to be 
\begin{equation}
    \gamma=\log 2+\log\left(1+\tanh^{2|h|}(\beta_A)\right). 
\end{equation}
We see that for $\beta_A\to\infty$ ($\Delta_A^2\to 0$), $\gamma=2\log 2$. However, for any finite $\beta_A$ (any nonzero $\Delta_A^2$), we get $\gamma=\log 2$ in the thermodynamic limit $|h|\to\infty$.

\subsection{Order and disorder parameters}\label{sorddisordAv}

Another way to show that (\ref{rhoAv}) describes a SW-SSB state is by computing its order and disorder parameters. (\ref{rhoAv}) has a strong (and therefore, also a weak) Wilson 1-form symmetry:
\begin{equation}
    \prod_{e\in \ell}Z_e\rho=\rho
\end{equation}
for any closed loop $\ell$. The strong symmetry is spontaneously broken because there is a nonzero order parameter:
\begin{equation}\label{strongzord}
    O_{2,x}=1,
\end{equation}
as defined in equation \eqref{eq:def_O2}.
In addition, the disorder parameter for the strong Wilson symmetry as defined in \eqref{eq:def_D1} is zero:
\begin{equation}\label{strongzdisord}
    D_{1,z}=0. 
\end{equation}
Equations (\ref{strongzord}) and (\ref{strongzdisord}) confirm that the strong Wilson symmetry is spontaneously broken. On the other hand, the weak Wilson symmetry is still present. The order parameter decays with an area law:
\begin{equation}
    O_{1,x}=\tanh^{|\ell|}(\beta_A)
\end{equation}
where $|\ell|$ is the area of the loop $\ell$. Furthermore, the disorder parameter is $\mathcal{O}(1)$:
\begin{equation}
    D_{2,z}=e^{-4\beta_A}. 
\end{equation}

Therefore, this density matrix demonstrates SW-SSB of the Wilson symmetry. There is also a weak 't Hooft 1-form symmetry:
\begin{equation}
    \prod_{e\in\ell}X_e\rho\prod_{e\in\ell}X_e=\rho
\end{equation}
This 1-form symmetry can be easily checked to be spontaneously broken: $O_{1,z}=1$ and the disorder parameter is zero.

\subsection{Choi state}\label{schoitemp}

We will now compute the Choi state for (\ref{rhoAv}) and show that it takes a particularly simple form, from which we can infer the TEE. It is convenient to first start with the mixed state 
\begin{equation}
    \rho_B=\frac{1}{2^{N-1}}\prod_p\left(\frac{1+B_p}{2}\right),
\end{equation}
Its Choi state $|\rho_B\rrangle$ takes a simple form:
\begin{align}
    |\rho_B\rrangle
    \propto\prod_e\left(\frac{1+Z_{e,u}Z_{e,l}}{2}\right)|tc_u\rangle|tc_l\rangle, 
\end{align}

We can apply the operator $\prod_v e^{-\beta_A A_v/2}$ on both sides of the density matrix $\rho_B$ to obtain our target density matrix (the correction due to exclusion of vertex $v_0$ will not affect the scaling behavior and will be dropped below)
\begin{equation}\label{rhobeta}
    \rho(\beta_A)\propto \prod_v e^{\beta_A A_v/2}\prod_p\left(\frac{1+B_p}{2}\right)\prod_{v'}e^{\beta_A A_{v'}/2}. 
\end{equation}
The Choi state for $\rho(\beta_A)$ then takes the following form
\begin{equation}
    |\rho(\beta_A)\rrangle\propto \prod_v e^{\beta_A A_{v,u}/2}\prod_{v'} e^{\beta_A A_{v',l}/2}|\rho_B\rrangle.
\end{equation}
It is convenient to introduce a new set of spins acted on by Pauli matrices $Z_{e,m},X_{e,m}$ (where $m$ refers to ``middle"), with the constraint 
\begin{equation}\label{zconstr}
    Z_{e,u}Z_{e,l}=Z_{e,m}. 
\end{equation}
We add $N_e$ spins, but imposing the $N_e$ constraints (\ref{zconstr}) takes reduces the Hilbert space to its original dimension. To preserve the commutation relations, we also need the maps
\begin{equation}\label{xconstr}
    X_{e,u}\to X_{e,u}X_{e,m},\qquad X_{e,l}\to X_{e,m}. 
\end{equation}

We will now work exclusively with spins in $\mathcal{H}_{u}$ and $\mathcal{H}_{m}$. This allows us to rewrite the state $|\rho_B\rrangle$ in a simple way:
\begin{equation}
    |\rho_B\rrangle=|tc_{u}\rangle|\uparrow_{m}\rangle,
\end{equation}
where $|\uparrow_{m}\rangle$ is a product state in the middle Hilbert space $\mathcal{H}_m$ with all spins pointing in the $+z$-direction. 
$|\rho(\beta_A)\rrangle$ takes the form
\begin{align}
\begin{split}
    |\rho(\beta_A)\rrangle&\propto\prod e^{\beta_A A_{v,u}A_{v,m}/2}\prod e^{\beta_A A_{v,m}/2}|tc_{u}\rangle|\uparrow_{m}\rangle\\
    &\propto|tc_{u}\rangle\prod e^{\beta_A A_{v,m}}|\uparrow_{m}\rangle.
\end{split}
\end{align}
Therefore, $|\rho(\beta_A)\rrangle$ is simply an unperturbed toric code state stacked with a confined state. 
The state in $\mathcal{H}_{m}$ is confined as easily evaluated from the correlation functions of loops. 
One can also ``ungauge" the $\mathbb{Z}_2$ symmetry and find that the state is connected to the ferromagnetic phase for any finite $\beta$. Specifically, the ungauged version of the state in $\mathcal{H}_m$ is
\begin{equation}
    \prod e^{\beta_A X_{v,m}}|GHZ_{m}\rangle
\end{equation}
The order parameter is $\mathcal{O}(1)$ for all finite $\beta$. Specifically, $\mathrm{Tr}(Z_iZ_j\rho)=e^{-4\beta_A}$, independent of $|i-j|$ for sufficiently separated $i,j$. Furthermore, the disorder parameter evaluates to (dropping the subscript $m$ in the following equation for ease of notation)

\begin{align}
\begin{split}
    &\frac{\langle GHZ|\prod e^{\beta X_v}\prod_{v\in A}X_v\prod e^{\beta X_v}|GHZ\rangle}{\langle GHZ|\prod e^{\beta X_v}\prod e^{\beta X_v}|GHZ\rangle}\\
    &=\frac{\langle GHZ|\prod_{v\in A}X_v\prod \left(\cosh(2\beta)+\sinh(2\beta)X_v\right)|GHZ\rangle}{\langle GHZ|\prod \left(\cosh(2\beta)+\sinh(2\beta)X_v\right)|GHZ\rangle}\\
    &=\frac{\cosh^{N-|A|}(2\beta)\sinh^{|A|}(2\beta)}{\cosh^{N}(2\beta)+\sinh^N(2\beta)}\\
    &=\frac{\tanh^{|A|}(2\beta)}{1+\tanh^N(2\beta)}\sim \tanh^{|A|}(2\beta)
\end{split}
\end{align}
where we used $\langle GHZ_{m}|X_{v,m}|GHZ_{m}\rangle=0$ and $\lim_{N\to\infty}\tanh^N(2\beta)=0$ for finite $\beta$. Since the order parameter is always $\mathcal{O}(1)$ and the disorder parameter always decays very quickly, with an area-law, we get a ferromagnetic Ising state for any finite $\beta$, which maps onto a confined state after gauging the $\mathbb{Z}_2$ symmetry. Strictly for $\beta=\infty$, the TEE is $2\log 2$, with $\log 2$ from each layer, while for any finite $\beta$, the TEE is $\log 2$ coming only from the spins in $\mathcal{H}_{u}$.

\subsection{Path of gapped parent Hamiltonians}\label{shamunitary}

To further support the results in the previous subsection, we show below that there is an explicit path of gapped, local Hamiltonians connecting a parent Hamiltonian for $|\rho(\beta_A)\rrangle$ to a parent Hamiltonian for $|\rho_B\rrangle$. This clearly demonstrates that these two states are in the same phase, and share the same TEE (modulo spurious contributions). The idea is to find an exact unitary operator $\mathcal{U}(\beta_A)_{m}$ which has the same action on $|\uparrow_{m}\rangle$ as $\prod e^{\beta_AA_{v,m}}$ (upon proper renormalization):
\begin{equation}
|\rho(\beta_A)\rrangle=\mathcal{U}_m(\beta_A)|tc_{u}\rangle|\uparrow_{m}\rangle
\end{equation}

This unitary operator takes a simple, exact form on a system with open boundary conditions, and we will show that it is locality-preserving for all local operators: it takes local operators to nearby local operators, up to exponential tails. It follows that 
\begin{equation}
    H(\beta_A)=\mathcal{U}_m(\beta_A)H_0 \mathcal{U}_m(\beta_A)^{\dagger},
\end{equation}
where $H_0$ is the parent Hamiltonian of $|tc_{u}\rangle|\uparrow_{m}\rangle$
\begin{equation}
    H_0=-\sum A_{v,u}-\sum B_{p,u}-\sum Z_{e,m}. 
\end{equation}
$H(\beta)$ is a local parent Hamiltonian for $|\rho(\beta_A)\rrangle$, whose terms are exponentially localized with a length scale set by $\beta_A$. 

In Appendix~\ref{ssequcirc}, we show that $U(\beta_A)_{m}$ can be written as the following sequential circuit:
\begin{align}
\begin{split}\label{seqcirc}
     &\mathcal{U}_m(\beta_A)\\
     &=\prod_{v_x;v_y=L-2} e^{i\tilde{\beta}_A\tilde{A}_{v,m}}\dots\prod_{v_x;v_y=1} e^{i\tilde{\beta}_A\tilde{A}_{v,m}}\prod_{v_x;v_y=0} e^{i\tilde{\beta}_A\tilde{A}_{v,m}}
\end{split}
\end{align}
where $v_x,v_y$ are $x$ and $y$ coordinates for a vertex and 
\begin{equation}
    \tan(\tilde{\beta}_A)=-\tanh(\beta_A)
\end{equation}

While sequential circuits can generally take one gapped ground state to a gapped ground state in a different phase \cite{Chen_2024}, the above sequential circuit maps \emph{all} local operators to nearby local operators and maintains the universal properties of the state. This is different from, for example, the sequential circuit that maps a paramagnet ground state to a GHZ state\cite{Chen_2024}.

Since the circuit (\ref{seqcirc}) is only sequential in the $y$ direction only, it leads to a finite correlation length only in the $y$ direction. In other words, $\mathcal{U}_m(\beta_A)O_{e,m}\mathcal{U}_m(\beta_A)^{\dagger}$ generically maps a local operator $O_{e,m}$ supported on site $e$ to an operator with strict locality in the $x$ direction but an exponential tail in the $+y$ direction (note that there is no tail in the $-y$ direction). In Appendix~\ref{ssequcirc}, we show that $\mathcal{U}(\beta_A)_{m}$ is locality-preserving for all local operators. In particular, for a verticle link $e$,
\begin{equation}\label{tildeZ}
\mathcal{U}_m(\beta_A)Z_{e,m}\mathcal{U}_m(\beta_A)^{\dagger}=\tilde{Z}_{e,m}.
\end{equation}
In the thermodynamic limit $L\to\infty$, we find that for an operator $O_{e,m}$ supported on a different $x$ coordinate commutes with $\tilde{Z}_{e,m}$. If it is supported on the same $x$ coordinate as $e$ with a different $y$ coordinate $y'>y$, then
\begin{equation}
    \|[\tilde{Z}_{e,m},O_{e,m}]\|\leq \mathrm{const}\cdot e^{-|y'-y|/\xi},
\end{equation}
where the correlation length $\xi$ in the $y$ direction is given by
\begin{equation}
    \xi=-\frac{1}{\log\left[|\sin(2\tilde{\beta}_A)|\right]}. 
\end{equation}
As expected, $\xi$ diverges as $\beta_A\to\infty$  ($\tilde{\beta}_A\to-\frac{\pi}{4})$ and goes to zero as $\beta_A\to 0$ 
 ($\tilde{\beta}_A\to 0)$. The spread of $\tilde{Z}_{e,m}$ where $e$ is a horizontal link can be handled similarly, and has the same correlation length. 

It follows from (\ref{tildeZ}) that
\begin{equation}
H(\beta_A)=-\sum A_{v,u}-\sum B_{p,u}-\sum\tilde{Z}_{e,m}
\end{equation}
is a gapped Hamiltonian that consists of exponentially local terms, as long as $\beta_A$ is finite. When $\beta_A=0$, the ground state of $H(\beta_A)$ is simply a toric code fixed point state stacked with a trivial paramagnet, so it clearly has TEE $\log 2$. The same must hold for the ground state $|\rho(\beta_A)\rrangle$ of $H(\beta_A)$ for any finite $\beta_A$.

\subsection{Random plaquette term}\label{srandplaquette}

The above discussion of Choi state, parent Hamiltonian, etc. can be generalized to the case where $\lambda_B$ also has exhibits randomness. If $\lambda_B$ also exhibits randomness, then the state is equivalent to the trivial infinite temperature state. By the same derivation as in Sec.~\ref{schoitemp}, we can map a disordered density matrix with random $\{\lambda_{B,p}\}$ drawn from a Gaussian distribution with mean 1 and variance $\Delta_B^2$ to a finite-temperature toric code model. This density matrix has the form
\begin{equation}
    \rho(\beta\lambda_A,\beta\lambda_B)\propto\prod e^{\beta \lambda_AA_v}\prod e^{\beta\lambda_B B_p}
\end{equation}
where for $\alpha=A, B$ we have
\be  \tanh(\beta\lambda_{\alpha})=\mathrm{erf}\left(\frac{1}{\sqrt{2}\Delta_{\alpha}}\right). 
\ee
Without loss of generality we can choose $\beta=1$, which then fixes $\lambda_A$ and $\lambda_B$ in terms of $\Delta_A$ and $\Delta_B$.

The corresponding Choi state can be written as
\be
    |\rho_{\beta}\rrangle\propto\prod_v e^{\beta\lambda_A (A_{v,u}+A_{v,l})/2}\prod_p e^{\beta\lambda_B (B_{p,u}+B_{p,l})/2}|\rho_1\rrangle, 
\ee
where $|\rho_1\rrangle$ is the Choi state of the identity matrix, which is a state where every spin in $\mathcal{H}_u$ is paired with a spin in $\mathcal{H}_l$ in a Bell pair. We can write $|\rho_1\rrangle$  as the state obtained from full decoherence of toric code with both the $X$ and $Z$ channels.
\begin{equation}
    |\rho_1\rrangle\propto\prod_{e,e'}\left(\frac{1+Z_{e,u}Z_{e,l}}{2}\right)\left(\frac{1+X_{e',u}X_{e',l}}{2}\right)|tc_{u}\rangle|tc_{l}\rangle. 
\end{equation}
As in Sec.~\ref{schoitemp}, we will rewrite the state with an additional set of spins acted on by $X_{m},Z_{m}$. Following (\ref{zconstr}) and (\ref{xconstr}), $|\rho_1\rrangle$ can be written as
\begin{equation}
    |\rho_1\rrangle=|+_u\rangle | \uparrow_m\rangle
\end{equation}

Therefore, $|\rho_{\beta}\rrangle$ can be written as
\begin{align}
\begin{split}
    |\rho_{\beta}\rrangle&\propto\prod_v e^{\beta \lambda_A (A_{v,u}A_{v,m}+A_{v,m})/2}\\
    &\times \prod_p e^{\beta\lambda_B (B_{p,u}B_{p,m}+B_{p,u})/2}
    |+_{u}\rangle|\uparrow_{m}\rangle\\
    &\propto\left(\prod_p e^{\beta \lambda_BB_{p,u}}|+_{u}\rangle\right)\left(\prod_v e^{\beta A_{v,m}}|\uparrow_{m}\rangle\right)\\
\end{split}
\end{align}
We can apply the same method as in Sec~\ref{shamunitary} to construct sequential circuits $\mathcal{U}_{u}$ and $\mathcal{U}_{m}$ such that
\begin{equation}
|\rho(\beta\lambda_A,\beta\lambda_B)\rrangle=\mathcal{U}_u(\beta\lambda_B)\mathcal{U}_m(\beta\lambda_A)|+_{u}\rangle|\uparrow_{m}\rangle
\end{equation}
It follows that 
\begin{align}
\begin{split}
    H(\beta\lambda_A,\beta\lambda_B)= & -\sum \mathcal{U}_u(\beta\lambda_B)X_{e,u}\mathcal{U}_u(\beta\lambda_B)^{\dagger}\\
    &-\sum \mathcal{U}_m(\beta\lambda_A)Z_{e,u}\mathcal{U}_m(\beta\lambda_A)^{\dagger}
\end{split}
\end{align}
is a gapped Hamiltonian with terms that decay as
\begin{equation}
    \xi_{u}=-{\log^{-1}\left[|\sin(2\tilde{\beta_B})|\right]},\quad \xi_{m}=-{\log^{-1}\left[|\sin(2\tilde{\beta_A})|\right]}
\end{equation}
where 
\begin{equation}
    \tan(\tilde{\beta}_{\alpha})=-\tanh(\beta\lambda_{\alpha}) 
\end{equation}
for $\alpha=A, B.$
Therefore, $H_{\beta}$ is local as long as $\beta\lambda_B$ and $\beta\lambda_A$ are both finite. In other words, as long as $\beta\lambda_B$ and $\beta\lambda_A$ are finite, there is a path of local, gapped Hamiltonians connecting a parent Hamiltonian for the trivial Choi state (the Choi state of the identity operator) to a parent Hamiltonian for the Choi state of $\rho(\beta\lambda_A,\beta\lambda_B)$.

\section{Example: random field}\label{sex2}

In this section, we will turn to a slightly more complicated disordered system. To the bare toric code Hamiltonian (\ref{tcham}), we add a field $Z_e$ on every link with random coefficients $\{h_e\}$:
\begin{equation}\label{hamrandfield}
    H(\{h_e\})=-\lambda_A\sum A_v-\lambda_B\sum B_p+\sum h_eZ_e. 
\end{equation}
Such a Hamiltonian is first studied in Ref. \cite{TC_field}. We will focus on the cases where distributions of $\{h_e\}$ have zero mean, as the resulting density matrix from summing over all disorder realizations will have an exact weak 't Hooft 1-form symmetry (see Appendix~\ref{sweakthooft}). We will start by examining the case where $\{h_e\}$ are taken from a bimodal distribution with zero mean, i.e. $h_e=\pm h$. Then we will further discuss the case where $\{h_e\}$ are taken from a Gaussian distribution with zero mean. An interesting observation is that some qualitative features depend on the distribution of the disorder. For Gaussian disorder, we show in Sec.~\ref{sgaussian} and Appendix~\ref{so1zgauss} that $O_{1,z}\sim e^{-\mu r/\xi}$ for all finite $\Delta_h$, for some constants $\mu,\xi$, even for $\lambda_B\ll \Delta_h$. On the other hand for bimodal disorder, $O_{1,z}\sim e^{-\mu r^2/\xi^2}$ for $\lambda_B\ll\Delta_h$ (we are sloppy with notation here; the constants are generally different from those in the Gaussian case). There are also distinct quantitative differences, arising from the fact that the ground state of $H(\{h_e\})$ at $\lambda_A=0$ is extensively degenerate for bimodal disorder and unique for Gaussian disorder \cite{fisher1987}.

We now focus on bimodal disorder with amplitude $|h_e|=h$. The ensemble of ground states of (\ref{hamrandfield}) tunes through the three different mixed phases described in Sec.~\ref{sphases}:
\begin{itemize}
    \item $h\ll\lambda_A,\lambda_B$: ST-SSB. The strong and weak Wilson and 't Hooft  1-form symmetries are both spontaneously broken. The Choi state describes two copies of toric code.
    \item $\lambda_A\ll h\ll\lambda_B$: SW-SSB. The strong Wilson 1-form symmetry is spontaneously broken but the weak Wilson 1-form symmetry is respected. There is only a weak 't Hooft 1-form symmetry, which is spontaneously broken. The Choi state describes a single copy of toric code.
    \item $\lambda_A,\lambda_B\ll h$: WS: There is only a weak Wilson symmetry and a weak 't Hooft symmetry, and they are both respected. The Choi state is trivial.  
\end{itemize}

In the following, we will discuss each of these phases in detail. In particular, we will use perturbation theory around the fixed points of the three phases to show that the phases occur over a finite window of $h$.

\subsection{ST-SSB phase}\label{sSt}

For $h\ll \lambda_A,\lambda_B$, the density matrix lies in the ST-SSB phase. We will show that if the random field is treated perturbatively, then $\rho(h)$ matches exactly with the density matrix of the decohered toric code with incoherent Pauli $Z$ errors, with 
\begin{equation}\label{phrelation}
    p\sim\frac{h^2}{16\lambda_A^2}, 
\end{equation}
where $p$ is the decoherence strength or the error rate. It is known that the decohered density matrix with $p$ sufficiently small is in the toric code phase; its Choi state is equivalent to two copies of toric code \cite{bao2023mixedstate} and it shares the same qualitative features as the pure toric code ground state. In particular, it has (emergent) spontaneously broken strong and weak Wilson 1-form symmetries \cite{mcgreevy2023generalized}. Therefore, using the map \eqref{phrelation} between the disordered and the decohered density matrices, we can leverage known results about the latter at small $p$ to describe the former at perturbatively small $h$.

We obtain the disordered ensemble $\rho(h)$ for $h\ll \lambda_A,\lambda_B$ using quasi-adiabatic continuation. Quasi-adiabatic continuation is a method for obtaining the ground state $|\psi(s)\rangle$ of a family of gapped Hamiltonians $H(s)$, parameterized by $s\in[0,1]$. The path of gapped Hamiltonians of interest in this case is
\begin{equation}
    H(s,\{h_e\})=-\lambda_A\sum A_v-\lambda_B\sum B_p-s\sum h_eZ_e. 
\end{equation}

We will assume that $h$ is small compared to $\lambda_{A/B}$, so the path of Hamiltonians from $s=0$ (the toric code fixed point Hamiltonian (\ref{tcham})) to $s=1$ (which is simply (\ref{hamrandfield})) remains gapped. As long as $H(s,\{h_e\})$ remains local and gapped, we can write the ground state along the path of gapped Hamiltonians as
\begin{equation}
    |\psi(s,\{h_e\})\rangle=\U (s,\{h_e\})|\psi(0,\{h_e\})\rangle
\end{equation}
where $\U(s,\{h_e\})$ is a unitary operator that can be written as finite time-evolution of a local Hamiltonian, and $|\psi(0,\{h_e\})\rangle=|tc\rangle$. In Appendix~\ref{squasi}, we show that in the limit $h\ll \lambda_A$, we can write $\U$ as
\begin{equation}\label{pertU}
    \U(s,\{h_e\})=e^{iDs},\quad D=\frac{i}{8\lambda_A}\sum_e h_eZ_e(A_v+A_{v'}).
\end{equation}
where $v,v'$ are the two vertices at the endpoints of the link $e$. 

Denoting $\U(s=1,\{h_e\}):=\U(\{h_e\})$ and $\psi(s=1,\{h_e\}):=\psi(\{h_e\})$, we can write the ground state projector as 
\begin{equation}
    |\psi(\{h_e\})\rangle\langle\psi(\{h_e\})|=\U(\{h_e\})|tc\rangle\langle tc|\U^\dagger(\{h_e\})
\end{equation}

Combining the two equations above, we arrive at the following density matrix for small $\frac{h}{4\lambda_A}$, 
\begin{equation}
    \rho\propto\prod\left(\frac{1+B_p}{2}\right)\sum_{\sigma}\prod_vA_v^{\sigma_v}\left(1-\frac{h^2}{16\lambda_A^2}\right)^{2N_{\sigma}}
\end{equation}
Here, we sum over products of $A_v$ operators labeled by $\sigma$. Specifically, each $\sigma$ is a vector with $N_v$ elements $\sigma_v=0,1$, and we sum over all such vectors. $N_{\sigma}$ is the total length of all the closed Pauli $X$ loops in a configuration labeled by $\sigma$. As we show in Appendix~\ref{squasi}, the density matrix above is identical to the decohered toric code with the relation (\ref{phrelation}).

The scaling of the order and disorder parameters follow exactly those of the decohered toric code at small amplitude of incoherent $Z_e$ noise (see Appendix~\ref{sec:decohered}):
\begin{align}
\begin{split}
    &O_{1,x}=e^{-2\mu r/\xi},\qquad O_{1,z}=1,\\
    &O_{2,x}=O_{2,z}=1,\\
    &D_{1,x}\sim D_{1,z}\sim e^{-\mu |i-j|/\xi},\\
    &D_{2,x}=0,\qquad D_{2,z}=e^{-\mu |i-j|/\xi}.\\
\end{split}
\end{align}
The first and third line follows from the fact that every state in $\rho$ is in the toric code phase. $D_{2,z}$ follows from the aforementioned mapping to the decohered toric code at small $p$.

\subsection{SW-SSB phase}\label{sSW}

Now we consider the regime $\lambda_A\ll h\ll\lambda_B$ and apply $\lambda_A$ perturbatively. At $\lambda_A=0$, the Hamiltonian is classical, and has an extensive number of exactly degenerate ground states\footnote{The degeneracy is absent if $\{h_e\}$ are taken from a Gaussian distribution \cite{fisher1987}.} (which we will discuss in more details in Sec.~\ref{sgaussian}). However, the $-\lambda_A\sum_V A_v$ term lifts the exact degeneracy, for general $\{h_e\}$ configurations, at extensively large order in degenerate perturbation theory. We use the resulting set of unique ground states to form the density matrix $\rho(\{h_e\})$. In this section, we will treat $-\lambda_A\sum A_v$ as a perturbation that only appears at the level of degenerate perturbation theory, and argue that the perturbed density matrix possesses similar order and disorder parameters as the ones in the random vertex example (\ref{rhoAv}) in the SW-SSB phase. Therefore, the density matrix must be in the SW-SSB phase of the Wilson 1-form symmetry as well.

The classical ground states of (\ref{hamrandfield}) with $\lambda_A=0$ can be reached by the following two-step process:
\begin{enumerate}
    \item Minimize all of the field terms $-\sum h_eZ_e$, i.e. $Z_e=\sgn(h_e)$. 
    \item Correct the $B_p=-1$ plaquettes by flipping the fewest possible spins and violating the fewest number of random field terms. 
\end{enumerate}

Step 2 is required as $\lambda_B\gg h$ is dominant  
and the ground state must satisfy $B_p=1$ for all plaquettes. While step 1 produces a unique state, there are generally many ways to perform step 2, each resulting in a different classical ground state. For example, if there are only two $B_p=-1$ plaquettes separated by a distance $n$ in the $x$ direction and $m$ in the $y$ direction, then they can be corrected by applying the shortest dual string of $X_e$ operators connecting the two plaquettes, and the number of violated field terms is given by the length of the string. However, there are
\begin{equation}\label{degbimodal}
    {n+m \choose n}={n+m \choose m}
\end{equation}
different paths of the same shortest length $n+m$. For this configuration of $\{h_e\}$, there is therefore a ${n+m \choose n}$-fold ground state degeneracy at $\lambda_A=0$. For example, if $n=m=1$, then there is a two-fold degeneracy of classical ground states $|\psi_1\rangle,|\psi_2\rangle$. A small $-\lambda_A\sum A_v$ term lifts this degeneracy at first order in ${\lambda_A/h}$, resulting in a unique ground state $\frac{1}{\sqrt{2}}(|\psi_1\rangle+|\psi_2\rangle)$ in a finite-sized system. Since a particular vertex term $A_v$ toggles between $|\psi_1\rangle$ and $|\psi_2\rangle$, the expectation value of this particular vertex term is 1 in this perturbed ground state. More generally, the ground state will be an equal weight superposition of all the ${n+m \choose n}$ degenerate ground states between these two plaquettes with $B_p=-1$. 

Now consider the entire ground state  $\ket{\psi(\{h_e\})}$ of a typical disorder realization $\{h_e\}$. The argument above shows that the ground state consists of many short strings to minimally correct the plaquette violations which leads to local coherence of $A_v$. In such a state, in order to have $ \bra{\psi(\{h_e\})}\prod_{v\in R}A_v \ket{\psi(\{h_e\})}\neq 0$ in a region $R$, we need to have two $B_p=-1$ plaquettes on  two diagonal corners of $R$ and no $B_p=-1$ plaquettes in any part of the region which will cut off the range of coherence of $A_v$. Since $\sgn \left(\prod_{e\in p}h_e\right)$ has probability $1/2$ of being either $1$ or $-1$ for every plaquette $p$, when averaging over all possible disorder realization, such a probability roughly decays as $\left(\frac{1}{2}\right)^{|R|}$.\footnote{There is also a subleading $\left(\frac{1}{2}\right)^{|\partial R|}$ term from the fact that even in such a configuration of $\{h_e\}$, $\bra{\psi(\{h_e\})}\prod_{v\in R}A_v \ket{\psi(\{h_e\})}\sim\left(\frac{1}{2}\right)^{2(n+m)}$ according to (\ref{degbimodal}).} One can see that the scaling of the order and disorder parameters match with those discussed in Sec.~\ref{sorddisordAv}. In particular, $O_{1,x}\sim \left(\frac{1}{2}\right)^{|R|}$.

\subsection{WS phase}\label{sWt}

In the limit $\lambda_A,\lambda_B\ll h$, we can treat $A_v,B_p$ as perturbations. The ground state $|\psi_0(\{h_e\})\rangle$ with $\lambda_A=\lambda_B=0$ is a randomly polarized state in the $Z$ basis. Applying perturbation theory on the randomly polarized state gives
\begin{equation}
    |\psi(\{h_e\})\rangle\propto|\psi_0(\{h_e\})\rangle+\frac{\lambda_A}{8h}\sum_v|\psi_v(\{h_e\})\rangle+\mathcal{O}\left(\left(\frac{\lambda}{h}\right)^2\right)
\end{equation}
where $|\psi_v(\{h_e\})\rangle=A_v|\psi_0(\{h_e\})\rangle$. Note that there is no $B_p$ contribution because $\langle\psi_i(\{h_e\})|B_p|\psi_0(\{h_e\})\rangle=0$ for any excited state $|\psi_i(\{h_e\})\rangle$.
The corresponding density matrix at lowest order in perturbation theory is
\be
    \rho(h)\propto\sum_{\{h_e\}}\rho_0(\{h_e\})\\
    +\frac{\lambda_A}{8h}\sum_{v,\{h_e\}}\{A_v,\rho_0(\{h_e\})\},
\ee
where $\rho_0(\{h_e\})=|\psi_0(\{h_e\})\rangle\langle\psi_0(\{h_e\})|$. We can simplify the above expression by using $\sum_{\{h_e\}}\rho_0(\{h_e\})\propto\mathbf{1}$ because it is simply a sum over projectors onto all product states in the $Z$ basis. Therefore, to lowest order in $\frac{\lambda_A}{h}$, we get
\begin{equation}\label{rhorp}
    \rho(h)=\frac{1}{2^{N_e}}\left(1+\frac{\lambda_A}{4h}\sum_vA_v\right)
\end{equation}

This density matrix exhibits the phenomenology of the randomly polarized phase, based on the behaviors of the order and disorder parameters: 
\begin{align}
\begin{split}
    &O_{1,x}=O_{1,z}=0,\\
    &O_{2,x}=O_{2,z}=1,\\
    &D_{2,x}=D_{2,z}=1.\\
\end{split}
\end{align}
The last line comes from 
\begin{align}
\mathrm{Tr}(\rho^2)=\frac{1}{2^{N_e}}\left(1+\frac{N_e\lambda_A^2}{16h^2}\right)
\end{align}
and 
\begin{equation}
  \mathrm{Tr}(\prod_{e\in \ell'}Z_e\rho \prod_{e\in \ell'}Z_e\rho)=\frac{1}{2^{N_e}}\left(1+\frac{(N_e-4)\lambda_A^2}{16h^2}\right)  ,
\end{equation}
where $\ell'$ is an open string. The above two equations give $D_{2,z}\sim 1-\frac{4}{N_e}$, which goes to 1 in the limit $N_e\to\infty$.

One can easily evaluate the reduced density matrix of $\rho$ for any region by tracing the $A_v$ operators outside of the region; the R\'enyi-2 CMI is zero. The density matrix is perturbatively close to the identity.

Note that we could have also approached this problem using quasi-adiabatic continuation, as in the ST-SSB regime. In Appendix~\ref{squasi}, we show that the quasi-adiabatic result matches with the results in this section, at lowest order in $\frac{\lambda_A}{4h}$. \footnote{By comparison, in the ST-SSB regime, quasi-adiabatic continuation is preferred over direct perturbation theory because the lowest order contribution to $\rho(h)$ from perturbation theory is at order $\frac{h^2}{\lambda_A^2}$, and going to second order in perturbation theory leads to terms that scale as $\frac{N_v h^2}{\lambda_A^2}$. In the WS regime, we do not encounter this subtlety because the lowest order contribution to $\rho(h)$ is $\frac{\lambda_A}{h}$.}

\subsection{Gaussian distribution}\label{sgaussian}

In the previous sections, we considered the model (\ref{hamrandfield}) with $\{h_e\}$ taken from a bimodal distribution centered about zero. We will now comment on the case where $\{h_e\}$ are taken from a Gaussian distribution with mean zero and variance $\Delta_h^2$. We still obtain the same three phases as described above for bimodal disorder. However, there are two major qualitative differences between bimodal disorder and Gaussian disorder:
\begin{itemize}
    \item For Gaussian disorder, there is always a finite probability for the random field terms $\{h_e\}$ to overpower $\lambda_B$ on some plaquettes, even for $\Delta_h\ll \lambda_B$. Similarly, even for $\Delta_h\gg\lambda_B$, there is a finite probability for $\lambda_B$ to overpower the $\{h_e\}$ terms.

    \item For Gaussian disorder, the $\lambda_A\to 0$ limit with $\lambda_B\gg \Delta_h$ has a unique ground state (see i.e. Ref.~\cite{fisher1987}). On the other hand, as discussed in Sec.~\ref{sSW}, for bimodal disorder the $\lambda_A\to 0$ limit with $\lambda_B\gg h$ is extensively degenerate. 
\end{itemize}

The first point above leads to the fact that for any finite $\Delta_h$,
\begin{equation}\label{gaussianperimeter}
    O_{1,z}\sim e^{-\mu r/\xi}. 
\end{equation}
Therefore, unlike with bimodal disorder, there is never an exact strong Wilson 1-form symmetry. However, for $\Delta_h\ll\lambda_B$, there is an emergent strong symmetry, at least perturbatively in $\frac{\Delta_h}{\lambda_B}$. The scaling (\ref{gaussianperimeter}) differs from the scaling in the bimodal case, for $\lambda_B\ll\Delta_h$. We derive (\ref{gaussianperimeter}) in Appendix~\ref{so1zgauss}, in the limit $\frac{\lambda_B}{\Delta_h}\ll 1$. 

Regarding the second point, there is a \emph{unique} ground state with Gaussian disorder because the random fields $\{h_e\}$ can be ordered by their magnitude. In the SW-SSB phase, one can perform non-degenerate perturbation theory on the unique ground state for every set of $\{h_e\}$, similar to how we approached the ST-SSB phase and the WS phase. Even though there is a unique ground state, there is an extensive number of exponentially close low energy states. However, these states differ from the ground state by an extensive Hamming distance \cite{10.1007/978-3-642-01957-9_7} and therefore do not mix with the ground state at first order perturbation theory.

In the limit $\lambda_B\to \infty$, we get
\begin{align}
\begin{split}
    \rho&=\frac{1}{2^{N_e}}\prod_p\left(1+B_p\right)\left(1+\frac{\lambda_A}{\Delta E_{\mathrm{loc}}}\sum_vA_v\right)\\
    &\propto\prod_p\left(\frac{1+B_p}{2}\right)\prod e^{\beta A_v}
\end{split}
\end{align}
where $\beta\sim\frac{\lambda_A}{\Delta E_{\mathrm{loc}}}$. $\Delta E_{\mathrm{loc}}$ is the local energy gap, which is $\sim\mathcal{O}(\Delta_h)$. Note that for any given disorder realization, there are some small patches of the system where perturbation theory does not work, because $h_e$ happens to be much smaller than $\lambda_A$. However these patches are rare and local, so the state can roughly be considered to be one with $h_e\gg\lambda_A$ everywhere except with some isolated small holes, which do not change global properties of the state.\footnote{A similar complication shows up when one tries to use quasiadiabatic continuation for Gaussian disorder like we did for bimodal disorder. Roughly speaking, we expect that quasiadiabatic continuation still works, but instead of expanding everything straightforwardly in small $\frac{\lambda_A}{h_e}$ to get a unitary that maps to the perturbed state like in Appendix~\ref{squasi}, we need to additionally act on the state with some local unitaries near the patches where $h_e\gg \lambda_A$ (a similar issue comes up for quasiadiabatic continuation for small $\frac{h_e}{\lambda_A}$). In the thermodynamic limit, there is always going to be large patches, but according to the Ising spin glass numerics discussed in Sec.~\ref{sising}, there should still be robust phases.} 

Clearly, the above density matrix also has CMI $\log 2$ and is in the SW-SSB phase like (\ref{rhoAv}). Note that there seems to be a distinct difference in the effective $\beta$ between the the ensemble with bimodal disorder and Gaussian disorder. 

For $\lambda_B\neq\infty$, the $\prod Z$ loop decays with a perimeter-law as shown in Appendix~\ref{so1zgauss}. Since $\lambda_B$ does not participate in the $A_v$ part of the density matrix, the $A_v$ part of the density matrix still takes the form $\propto\prod e^{\beta A_v}$ with $\beta\sim\frac{\lambda_A}{\Delta E_{\mathrm{loc}}}$. The order and disorder parameters are given by
\begin{align}
\begin{split}
    &O_{1,x}=\tanh^{r^2}(\beta_A) \qquad O_{1,z}=e^{-\mu r/\xi}\\
    &O_{2,x}=O_{2,z}=1\\
    &D_{1,x}\sim e^{-\mu\mathrm{supp}(D_1)/\xi}\qquad D_{1,z}\sim e^{-\mu|i-j|/\xi}\\
    &D_{2,x}=e^{-\mu|i-j|/\xi}\qquad D_{2,z}=e^{-4\beta_A}\\
\end{split}
\end{align}
which match with the expected order and disorder parameters of the SW-SSB phase (here $r$ always refers to linear distance).

\subsection{Relation to Ising spin glass via gauging}\label{sising}

The toric code with random field is related to the Ising model with random coupling via gauging. Specifically, the disordered Hamiltonian in Eq. \ref{hamrandfield} at $\lambda_B\gg \Delta_h$ can be mapped to a 2$D$ random-bond Ising model in transverse field \cite{TC_field}:
\begin{equation}
    H=-\lambda_A\sum \sigma^x_v-\sum h_{e(vv')}\sigma^z_v\sigma^z_{v'}
\end{equation}
where $e(vv')$ labels the edge with endpoints at sites $v, v'$, and local exchange $\{h_e\}$ are taken from a probability distribution (i.e. bimodal or Gaussian) with mean zero. Writing explicitly, the gauging map from the Ising model to the toric code state reads
\begin{equation}
    \sigma^z_v\sigma^z_{v'}\to Z_{e(vv')},\ \sigma^x_v\to A_v.
\end{equation}
Extensive numerical studies have shown that at zero temperature and $\{h_e\}$ being drawn from Gaussian distributions, increasing the transverse field $\lambda_A$ drives a transition between a paramagnet phase and a spin glass phase, with the critical point at $\frac{\lambda_A}{\Delta_h}\sim 0.608$ \cite{fedorov1986quantum, rieger1994zero,guo1994quantum,thill1995equilibrium,read1995landau,Huse}. The spin glass phase is diagnosed by the non-vanishing Edwards-Anderson (EA) order parameter
\begin{equation}\label{EA}
    \tilde{O}_\text{EA}=\overline{|\langle \sigma^z_i\sigma^z_j\rangle|^2}=\sum p_n\left\vert\langle\psi_n|\sigma^z_i\sigma^z_j|\psi_n\rangle\right\vert^2,
\end{equation}
where $\langle\cdots\rangle$ denotes the expectation with respect to the ground state  $\ket{\psi_n}$ for each disorder realization $n$, and $\overline{\cdots}$ denotes the disorder average. On the other hand, the R\'enyi-2 disorder parameter for the toric code state,
\begin{equation}
    D_{2}(i,j)=\frac{\tr\left(\rho\prod_{e\in\gamma_{ij} }Z_e\rho\prod_{e\in\gamma_{ij}} Z_e\right)}{\tr\left(\rho^2\right)},
\end{equation}
is mapped to R\'enyi-2 correlator of the Ising spins
\begin{equation} \label{isingRényi2}
    \tilde{O}_{2}=\frac{\tr\left(\tilde{\rho}\sigma^z_i\sigma^z_j\tilde{\rho}\sigma^z_i\sigma^z_j\right)}{\tr\left(\tilde{\rho}^2\right)}=\frac{\sum_n p_n^2\left\vert\bra{\psi_n}\sigma^z_i\sigma^z_j\ket{\psi_n}\right\vert^2}{\sum_np_n^2},
\end{equation}
where $\tilde{\rho}=\sum_np_n\ket{\psi_n}\bra{\psi_n}$ denotes the density matrix for the disordered Ising model, and we assume that the ground states between different disorder realizations satisfy $\bra{\psi_m}\sigma^z_i\sigma^z_j\ket{\psi_n}=\bra{\psi_m}\sigma^z_i\sigma^z_j\ket{\psi_m}\delta_{m,n}$. The two order parameters $\tilde{O}_\text{EA}$ and $\tilde{O}_2$ do not exactly match, but they signal the same phase transition of the SW-SSB of the $Z_2$ global symmetry with slightly different critical point and critical behavior \cite{lessa2024strongtoweak}.\footnote{ In fact, as pointed out by Ref. \cite{lessa2024strongtoweak}, the fidelity correlator $\tilde{O}_F=\sqrt{F}\left(\tilde{\rho},\sigma^z_i\sigma^z_j\tilde{\rho}\sigma^z_i\sigma^z_j\right)$ will reduce to the EA order parameter once we assume the same condition of the density matrix as in the second equality in Eq. \ref{isingRényi2}. Here the fidelity is $\sqrt{F}(\rho,\sigma)\equiv \tr\left(\sqrt{\sqrt{\rho}\sigma\sqrt{\rho}}\right)$.} Therefore, we can view the transition from the toric code state to the SW-SSB phase of the Wilson 1-form symmetry is dual to the paramagnet-spin glass phase transition after gauging the global $\Z_2$ symmetry. Here, the SW-SSB phase of the global $\Z_2$ symmetry is mapped to the SW-SSB phase of the Wilson 1-form symmetry. Such a duality map can be straightforwardly generalized to higher dimensions: under the gauging map, SW-SSB phase of a 0-form global symmetry in $d+1$ dimensional spacetime is mapped to an SW-SSB state of its dual $(d-1)$-form  Wilson symmetry.

The plaquette term $B_p$ in the toric code is absent (or trivially satisfied) in the Ising Hilbert space. However, for any fixed flux pattern with $B_p=-1$ for some plaquettes, one can redefine the spins to absorb the signs. Alternatively, one can enlarge the Hilbert space by introducing symmetry defects, namely $\pi$ fluxes, of the global $\Z_2$ symmetry in the Ising model.

\section{Discussion}\label{sdiscussion}

In this work, we introduced an equivalence relation on density matrices based on finite-time local Lindbladian evolution maintaining finite R\'enyi-2 Markov length. One benefit of using this definition as opposed to the usual definition based on two-way channel connectivity is that it distinguishes all ``intrinsically mixed" phases from phases equivalent to pure states, including those like the classical memory state (\ref{bp}). These intrinsically mixed phases demonstrate spontaneously broken strong 1-form symmetries that are not compatible with those observed in pure gapped ground states; they describe non-modular anyon theories. They can also be understood as SW-SSB phases, where a non-modular 1-form symmetry demonstrates SW-SSB. Such phases can be found naturally in ensembles of ground states of disordered toric code Hamiltonians, and are robust in the sense that they exist over a finite parameter range, according to our equivalence relation.

\begin{figure}[htbp]
\includegraphics[scale=0.25]{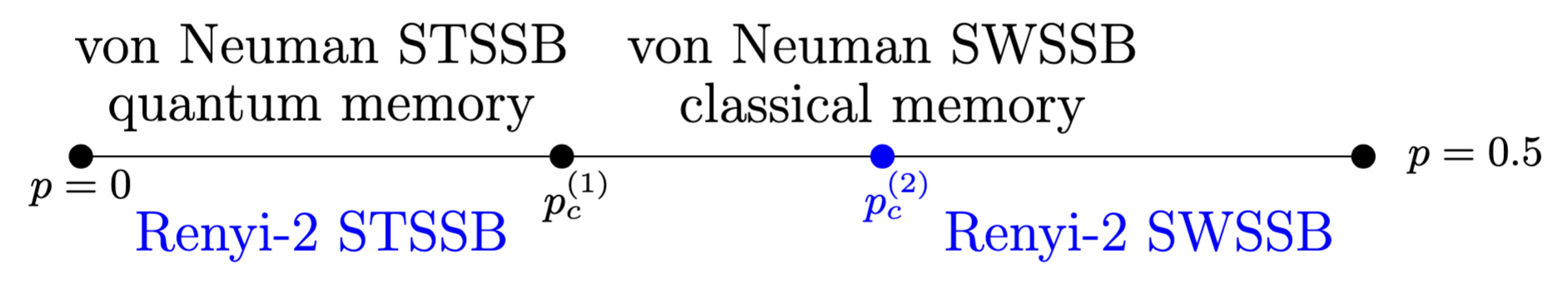}
\caption{The phase diagram of toric code with incoherent Pauli $X$ or $Z$ noise (\ref{Eq: Z dephasing}).}
\label{fig:decohered}
\end{figure}

We remark that our definition of mixed-state phases using R\'enyi-2 Markov length in general does not coincide with the definition using von Neumann Markov length \cite{sang2024Markov},  fidelity \cite{lessa2024strongtoweak} or Wightman Correlators \cite{liu2024diagnosingstrongtoweaksymmetrybreaking}. Our definition leads to a finer classification of phases than the usual definition of two-way channel connectivity and is complementary to existing methods in the sense that it inserts more phase boundaries in the phase diagram in a consistent way. In the context of decohered toric code, for example, 
our critical error rate $p_{c}^{(2)}$ is known to be greater than the $p_c^{(1)}$ where topological memory is lost. We chose Rényi-2 quantities because they lead to a clear relation with the Choi states. For decohered toric code detailed in appendix \ref{sec:decohered}, we expect that when $p>p_c^{(2)}=0.178$ the system is in the strong-to-weak SSB phase, serving as a classical memory, with Rényi-2 conditional mutual information $\log 2$. When $p<p_c^{(1)}=0.109$, the weak decoherence leaves the topological quantum memory intact, the system is in strong-to-trivial SSB phase and CMI is $2\log 2$ with the factor of two coming from two copies of toric code in the Choi state. In between $p_c^{(1)}<p<p_c^{(2)}$, it is two-way channel connected to $p=\frac{1}{2}$ but the corresponding Lindbladian evolution does not leave the Rényi-2 Markov length finite. While we expect this region to show no sign of Rényi-2 SWSSB, it exhibits nontrivial fidelity and Wightman correlator, indicating ``von Neumann" SWSSB. See the schematic phase diagram in Fig. \ref{fig:decohered}.

For simplicity, we considered just two particular kinds of disorder (random $A_v$ term and random $Z_e$ field). However, the perturbative methods introduced in Sec.~\ref{sex2} based on averaging over states obtained by quasi-adiabatic continuation should generalize broadly to other kinds of disorder. For example, we expect that we can apply similar methods to deal with Hamiltonians that also have a small $-\sum_eh_{e} X_e$ term. This term explicitly breaks the exact strong Wilson symmetry, but we expect to observe an emergent strong Wilson symmetry for sufficiently small disorder amplitude.

We only studied simple models related to the toric code. Models related to more complicated topological orders provide a rich playground for exploring other kinds of mixed states. 
For instance, imTOs examined in section II.D of reference \cite{ellison2024classification} in the context of decohered topological phases can also be obtained from disordered setup similar to ours, and be characterized by their patterns of SSB. As an example, one can add bimodal random fields to $\Z_4$ toric code which locally create $e^2m^2$ excitations. At intermediate disorder strength, we expect the $e^2m^2$ 1-form symmetry to exhibit SW-SSB, $e^2$ and $em$ symmetries, which commute with $e^2m^2$, to have ST-SSB, and $e$, $m$ to show WT-SSB. This $\mathbb{Z}_4$ example is also an example where even though there is SW-SSB, the state is conjectured to not be two-way channel connected to a pure state, because the transparent boson $e^2m^2$ is generated by the non-transparent line $em$. In other words, it does not factorize into $\mathcal{C}\boxtimes\mathcal{T}$ where $\mathcal{C}$ is modular and $\mathcal{T}$ contains only transparent bosons \cite{ellison2024classification}.
    
We caution against generalizing from the toric code example and assuming that all topological orders with strong decoherence, when using short anyon string operators, will result in SW-SSB; this is not always true. For example, we can consider the doubled semion topological order, with incoherent noise described by the following channel:
\begin{equation}
    \rho\to (1-p)\rho+pW_e\rho W_e^\dagger
\end{equation}
where $W_e$ is a short string operator for the semion (we can also consider the doubled semion Hamiltonian with random $h_eW_e$ terms). At $p=\frac{1}{2}$, we actually do not get a SW-SSB/intrinsically mixed phase, like we do when $W_e$ is a short string operator $Z_e$ in toric code. This can be seen by examining the Choi state. The Choi state for the initial pure doubled semion state describes two copies of doubled semion topological order: $|\rho\rrangle=|ds\rangle|ds\rangle$. The above channel at $p=\frac{1}{2}$ condenses $s_u\bar{s}_l$ in the Choi state, where $s_u$ is the semion of the upper layer and $\bar{s}_l$ is the antisemion of the lower layer. Therefore, the state after $p=\frac{1}{2}$ decoherence is symmetric under the $s_u\bar{s}_l$ weak 1-form symmetry. The remaining strong symmetries are string operators for $\bar{s}_u$ and $s_l$, which demonstrate ST-SSB (because the string operator of $\bar{s}_1s_2$ is also SSB). Therefore, we just have ST-SSB rather than SW-SSB, and a WS 1-form symmetry rather than one that is WT-SSB. It is easy to see that the Choi state describes a single copy of doubled semion at $p=\frac{1}{2}$; the mixed density matrix is equivalent to that of a chiral semion topological order, as discussed in Ref.~\cite{sohal2024noisy}. 

More generally, we conjecture that 1-form SW-SSB can only occur if the strong symmetries are not modular (for example, if there are transparent bosons), meaning SW-SSB phases coincide with intrinsically mixed phases. This is because if a strong symmetry, generated by $W_u$ in the Choi state, is spontaneously broken and its weak version $W_uW_l^\dagger$ is not, then $W_u$ has to braid trivially with $W_uW_l^\dagger$. Otherwise, it would serve as an order parameter for $W_uW_l^\dagger$. We leave a more detailed study of these phases and the contexts in which they can arise, such as decoherence and disorder, to future work.

Another important direction for future work is a clearer understanding of the relationship between equivalence relations on density matrices and equivalence relations on gapped Lindbladians. In isolated quantum systems, we have an equivalence relation on gapped ground states: two gapped ground states are in the same phase if they are related by finite time evolution of a local Hamiltonian. We also have an equivalence relation on gapped Hamiltonians: two gapped Hamiltonians are describe the same phase if there is an interpolation between them that maintains the gap. Finally, we also know that equivalence of parent Hamiltonians imply equivalence of their ground states, and vice versa. In particular, there is a way to get a unitary operator relating two gapped ground states if there is a gapped path between their parent Hamiltonians. 

For open quantum systems, as we discussed in this work, there are different equivalence relations on density matrices. One can also study equivalence relations for gapped Lindbladians, which has such density matrices as their steady states. There has been some work in this direction, but a complete story, and especially a relation to equivalence of steady states, is lacking. For example, it would be especially useful to obtain an effective action on steady states, i.e. a map from $\rho$ to $\rho'$, given a path of gapped Lindbladians from $\mathcal{L}_1$ to $\mathcal{L}_2$, similar to the unitary produced by quasiadiabatic evolution.

Finally, a subtle point about our equivalence relation is that we require two equivalent states to be connected to each other by finite time Lindbladian evolution with analytically  varying, finite R\'{e}nyi-2 Markov length. However, the ``analytically varying" requirement may not be necessary. 
If this were not true for von Neumann Markov length, and the Markov length can change discontinuously, then it seems that it may be possible to construct a path between the toric code and the trivial state maintaining finite, but discontinuously jumping Markov length, similar to a first order transition. However, this would mean that the toric code can also be recovered from the trivial state, because a discontinuity in the Markov length does not preclude a quasi-local recovery channel. Therefore, it seems that at least in some cases, it should not be possible to make at least the von Neumann Markov length jump discontinuously with local Lindbladian evolution. We leave a more in-depth analysis of this conjecture for future work.

\section*{Acknowledgements}
We thank Yu-Hsueh Chen, Meng Cheng, Tarun Grover, Timothy Hsieh, Chao-Ming Jian, Curt von Keyserlingk, Shang Liu, Ruochen Ma, Abhinav Prem, Tomohiro Soejima, Senthil Todadri, Ruben Verresen, Ashvin Vishwanath, and Chong Wang for inspiring discussions. We especially thank Rahul Sahay for calculations related to Appendix~\ref{shamunitary} and Shengqi Sang for many helpful discussions related to Markov length. Zhu-Xi thanks Jong Yeon Lee and Subir Sachdev for discussion on the replica trick. YX thanks Chao-Ming Jian for collaborating on related works. This research is partially supported by the Simons Collaboration on Ultra Quantum Matter which is a grant from the Simons Foundation (651440, Z.-X. L.), and grant NSF PHY-2309135 to the Kavli Institute for Theoretical Physics (KITP). YX acknowledges support
from the NSF through OAC-2118310. JHZ is supported by the U.S. Department of Energy under the Award Number DE-SC0024324. C.X. is supported by the Simons Foundation through the Simons Investigator program. 
\appendix

\section{Two-way channel between completely mixed state and SW-SSB}
\label{app:channel}

We will show in this appendix that there is an explicit two-way finite depth quantum channel between the maximally mixed state and the SW-SSB state $\rho_B=\frac{2^{N_p}}{2^{N_e}}\prod_p\left(\frac{1+B_p}{2}\right)$ on a sphere. Getting from $\rho_B$ to the maximally mixed state is straightforward -- one can simply apply the incoherent $X$ error channel, which is finite depth. To get from maximally mixed state to $\rho_B$, we first apply a complete amplitude damping channel to map to the product state $|\uparrow\rangle\langle\uparrow|$. It is instructive to first construct the complete amplitude damping channel for a single qubit. We want to map
\begin{equation}
|\downarrow_s\rangle|\downarrow_a\rangle\to|\downarrow_s\rangle|\downarrow_a\rangle\qquad|\uparrow_s\rangle|\downarrow_a\rangle\to|\downarrow_s\rangle|\uparrow_a\rangle,
\end{equation}
where $s$ stands for the system of interest, and $a$ stands for ancilla. 

This can be implemented by the unitary operator
\begin{equation}\label{Uspinancilla}
    U=X_aX_s\left(\frac{1-Z_aZ_s}{2}\right)+\left(\frac{1+Z_aZ_s}{2}\right).
\end{equation}
Generalizing to a bigger system size, we can use the finite depth unitary $\prod_eU_e$ where each $U_e$ takes the form (\ref{Uspinancilla}) between a spin $s_e$ and its ancilla $a_ee$. The complete amplitude damping channel is then given by
\begin{equation}
\mathrm{Tr}_a\left(\prod_eU_e\left(\frac{1}{2^{N_e}}\mathbf{1}\otimes|\downarrow_a\rangle\langle\downarrow_a|\right)\prod_eU_e^\dagger\right)=|\downarrow_s\rangle\langle\downarrow_s|
\end{equation}
where by abuse of notation, $|\downarrow_a\rangle\langle\downarrow_a|$ and $|\downarrow_s\rangle\langle\downarrow_s|$ are product states in the ancilla and spin spaces (not just single spin states).

Now we can apply another finite depth quantum channel to get from the product state $|\downarrow_s\rangle\langle\downarrow_s|$ to $\rho_B$. The initial state is
\begin{equation}
|\downarrow_s\rangle\langle\downarrow_s|=\prod_e\left(\frac{1-Z_e}{2}\right). 
\end{equation}
We now insert ancilla degrees of freedom on the vertices of the square lattice, in a product state in the $Z$ basis, and apply the unitary that maps 
\begin{equation}
    U_{cl}:Z_e\to X_vZ_eX_{v'}
\end{equation}
where $X_v$ and $X_{v'}$ act on the endpoints of the edge $e$. The unitary that performs this action is just the cluster state entangler except with a rotation between $X$ and $Z$, i.e. $U_{cl}=\prod e^{\frac{i\pi}{4}\sum X_eX_v}$. It also maps $Z_v\to Z_vA_v$. After tracing over the ancilla (vertex) degrees of freedom, only the products of $Z_e$ around plaquettes survive, because for such operator the $X_v$ operators on the vertices cancel. Therefore, the following finite depth channel takes the product state to $\rho_B$: 
\begin{equation}
    \mathrm{Tr}_a(U_{cl}|\downarrow_s\rangle\langle\downarrow_s|\otimes|\downarrow_a\rangle\langle\downarrow_a|U_{cl}^\dagger)=\rho_B
\end{equation}

On the torus, this finite-depth channel produces a state that also has two global projectors $\frac{1+\prod_e Z_e}{2}$ along the two cycles of the torus. This is because these operators are products of local operators $X_vZ_eX_{v'}$ that also survive the trace; all of the vertex terms cancel. This is the analogue of a short range entangled basis for the SW-SSB ``ground state space," like the fully polarized basis for ground state space of the $\mathbb{Z}_2$ SSB phase. A state on the torus without those two global projectors would have nontrivial long range connected correlation functions of non-contractible loops $\mathrm{Tr}(W_{x}W_{x}'\rho)-\mathrm{Tr}(W_x\rho)\mathrm{Tr}(W_{x}'\rho)=1$ where $W_x$, as discussed further in Sec.~\ref{sec:degeneracy}, is a non-contractible loop winding the $x$ cycle (and $W_x'$ is a similar loop but at a shifted position). Such a state would not be preparable by a finite-depth circuit from the fully mixed state. It is similar to a cat state, but rather than a coherent superposition, it is an incoherent superposition. In particular, it can be formed as a sum of two states preparable from the fully mixed state, with different loop parities along the $x$ cycle.

Note that the finite depth quantum channel above breaks the Wilson 1-form symmetry explicitly (in addition, the channel taking $\rho_B$ to maximally mixed state is also not 1-form symmetric). If we stay within the space of finite depth 1-form symmetric channels, then the two states are not two-way connected. Another subtlety is that the above channels are finite depth, but they can only be approximated by Lindbladian evolution with $\log L$ time, where $L$ is the system size (see i.e. Ref.~\cite{sang2024Markov} and Appendix~\ref{slogtime}).

\section{Choi state TEE and LW CMI}\label{sentanglement}

In this section we will directly compute the Choi state TEE and density matrix LW CMI at the fixed points described in Sec.~\ref{sphases}. We hypothesize that these quantities match even away from the fixed points. We showed that this holds at least for the density matrices described in Sec.~\ref{sex1}. 

The Choi state in all three phases described in Sec.~\ref{sphases} (ST-SSB, SW-SSB, and WS) is gapped, and has area-law entanglement entropy. We can isolate the subleading $\mathcal{O}(1)$ term in the entanglement entropy, which is the TEE, using the Levin-Wen subtraction scheme; the result is simply the CMI with regions $A,B,C$ chosen as shown in Fig.~\ref{fig:lw}. Note that the entanglement entropy of the Choi state of an operator is also called the \emph{operator entanglement} \cite{Bertini_2020}. 
Therefore, the Choi state TEE is the topological operator entanglement of the density matrix. 

We first mention the TEE of the fixed point Choi states, which can be read off from the corresponding topological order. The TEE is $2\log 2$ in the ST-SSB phase because the Choi state is described by two copies of Toric code. In the SW-SSB phase, the Choi state TEE is $\log 2$. Finally in the WS phase, the Choi state TEE is zero. We will now show that in the ST-SSB phase, the LW CMI of the density matrix is $2\log 2$ while in the SW-SSB phase, the LW CMI is $\log 2$. Finally in the trivial phase, the LW CMI is zero. 

In fact, in Sec.~\ref{sex1}, we already computed the CMI and showed that it is $2\log 2$ at the ST-SSB fixed point and $\log 2$ at the SW-SSB fixed point. At the WS fixed point, the density matrix is proportional to the identity operator. Therefore, the LW CMI is clearly zero. 

We will now present a different way to compute the R\'enyi-2 LW CMI. We present this different approach because it might allow for a direct derivation that the LW CMI and Choi state TEE must match. 

Let us denote the complement of $A\cup B\cup C$ in Fig.~\ref{fig:lw} by $D$. Then by plugging (\ref{Rényi2choi}) into (\ref{Rényi2cmi}) with the partitions in Fig.~\ref{fig:lw}, we get
\begin{align}
    \begin{split}\label{gammarho0}
&\gamma(\rho)\\
&=-\log\frac{\sum_{\sigma_n\in C\cup D,\sigma_m\in A\cup D}\llangle\rho|\sigma_{n,u}\sigma_{n,l}^\dagger|\rho\rrangle\llangle\rho|\sigma_{m,u}\sigma_{m,l}^\dagger|\rho\rrangle}{\sum_{\sigma_n\in A\cup C\cup D,\sigma_m\in D}\llangle\rho|\sigma_{n,u}\sigma_{n,l}^\dagger|\rho\rrangle\llangle\rho|\sigma_{m,u}\sigma_{m,l}^\dagger|\rho\rrangle}
\end{split}
\end{align}

Like in (\ref{I2choi}), we find that $\gamma(\rho)$ can be written in terms of expectation values of operators in the Choi state. Let us first evaluate $\gamma(\rho)$ when $|\rho\rrangle=|tc\rangle|tc\rangle$ (i.e. $\rho=|tc\rangle\langle tc|$), where $|tc\rangle$ is the fixed point toric code wavefunction. 
In this case, the only nonzero expectation values are those of closed $X_uX_l$ and $Z_uZ_e$ loops. These operators have expectation value 1. The number of closed $X_uX_l$ loops in a simply connected region $R$ is $2^{N_{v,R}}$ while the number of $Z_uZ_e$ loops is $2^{N_{p,R}}$. It follows that
\begin{equation}\label{gammarhotc}
\gamma(\rho)=-\log\frac{(2^{N_{v,C\cup D}}2^{N_{p,C\cup D}})(2^{N_{v,A\cup D}}2^{N_{p,A\cup D}})}{(2^{N_{v,A\cup C\cup D}}2^{N_{p,A\cup C\cup D}}\times 4)(2^{N_{v,D}}2^{N_{p,D}})}
\end{equation}
Note that there is an extra factor of four from the set of operators supported in $A\cup C\cup D$, which is not simply connected. The extra factor comes from $X_uX_l$ and $Z_uZ_e$ loops around either the left or the right subset of $B$; these loops also have expectation value 1 but are not generated by local loops fully supported in $A\cup C\cup D$. All of the other factors cancel between the numerator and denominator above, so we get
\begin{equation}
    \gamma(\rho)=2\log 2. 
\end{equation}

We can also compute $\gamma(\rho)$ easily in the SW-SSB phase. For the fixed point state $|\rho\rrangle\llangle\rho|=4\prod_e\left(\frac{1+Z_{e,u}Z_{e,l}}{2}\right)\prod_v \left(\frac{1+A_{v,u}A_{v,l}}{2}\right)\prod_p\left(\frac{1+B_{p,u}}{2}\right)$, the only operators with nonzero expectation value are products (not necessarily closed loops) of $Z_{e,u}Z_{e,l}$ and closed $X_{e,u}X_{e,l}$ loops. We find that
\begin{equation}\label{gammarho}
\gamma(\rho)=-\log\frac{(2^{N_{e,C\cup D}}2^{N_{v,C\cup D}})(2^{N_{e,A\cup D}}2^{N_{v,A\cup D}})}{(2^{N_{e,A\cup C\cup D}}2^{N_{v,A\cup C\cup D}}\times 2)(2^{N_{e,D}}2^{N_{v,D}})}. 
\end{equation}
Note that we only get one factor of two from operators supported in $A\cup C\cup D$ because we only have the $X_{e,u}X_{e,l}$ non-contractible loop around half of $B$; the $Z_{e,u}Z_{e,l}$ loop is already built out of the $2^{N_{e,A\cup C\cup D}}$ operators with nonzero expectation value. Therefore, we get $\gamma(\rho)=\log 2$. A similar calculation shows that the trivial memory fixed point has $\gamma(\rho)=0$.

\section{R\'enyi-2 local indistinguishable sectors and SW-SSB}
\label{sec:degeneracy}

In this appendix, we will discuss how SW-SSB can give rise to R\'enyi-2 locally indistinguishable states on a torus. We consider the simplest case, putting the state (\ref{SW-SSBz2}) on a torus. We will show in this section that this SSB pattern leads to four locally R\'enyi-2 indistinguishable states on a torus, which is related to the fact that the Choi state describes a single copy of toric code. Other SW-SSB/intrinsically mixed phases demonstrate similar properties. We will restrict to the case where the symmetries are exact, but one can generalize the discussion in this appendix to the case where the symmetries are only emergent. The non-contractible loop operators would have to be dressed as discussed in Sec.~\ref{semergent}.

On the torus, we can define $W_{x}, W_y, T_x,$ and $T_y$, which are the $\mathbb{Z}_2$ Wilson and 't-Hooft 1-form symmetry operators along non-contractible cycles in the $x/y$ directions. It is well known that in an isolated quantum system, the toric code Hamiltonian has four locally indistinguishable ground states on a torus, coming from complete spontaneous breaking of this $\mathbb{Z}_2\times\mathbb{Z}_2$ 1-form symmetry. In particular, at the toric code fixed point, all four non-contractible loop operators commute with the Hamiltonian. We can choose a basis for the ground state space that consists of eigenstates of $W_x$ and $W_y$, so $W_x|\psi_1\rangle=\alpha_x|\psi_1\rangle$, $W_y|\psi_1\rangle=\alpha_y|\psi_1\rangle$. However, due to the commutation relations between $W_y$ and $T_x$, there is another state $T_y|\psi_1\rangle=|\psi_2\rangle$ that has eigenvalue $-1$ under $W_x$: $W_x(T_y|\psi_1\rangle)=-\alpha T_y|\psi_1\rangle=-\alpha|\psi_2\rangle$. Similarly, we can apply $T_y$ and $T_xT_y$ on $|\psi_1\rangle$ to get two other states orthogonal to $|\psi_1\rangle$. In total, we get four orthogonal states on a torus, labeled by their different eigenvalues under $W_x$ and $W_y$: $\{|\psi_i\rangle\}=\{|\psi_1\rangle,T_x|\psi_1\rangle,T_y|\psi_1\rangle,T_xT_y|\psi_1\rangle\}$. These states form a basis for the ground state space of the toric code Hamiltonian. Importantly, the above four states are locally indistinguishable:
\begin{equation}\label{indpure}
    \langle\psi_i|O|\psi_j\rangle=c_{O}\delta_{ij}
\end{equation}
for any local operator $O$.\footnote{Clearly $\langle\psi_i|T_x|\psi_j\rangle\neq c_{T_x}\delta_{ij}$ because $T_x$ is a global operator.} Here, $c_{O}$ is a constant: the expectation value of any local operator is the same in each of the ground states. The local indistinguishability of the toric code ground states is what allows them to form a robust quantum code.

For a density matrix with a strong $\mathbb{Z}_2$ Wilson symmetry that is SW-SSB and a weak $\mathbb{Z}_2$ 't Hooft symmetry demonstrating WT-SSB, there is a similar story: the SSB pattern leads to four R\'enyi-2 locally indistinguishable density matrices on a torus. We again label the non-contractible loop operators by $W_x,W_y,T_x,$ and $T_y$. If we have $W_x\rho_1=\alpha_x\rho_1$, then there is another density matrix $T_y\rho_1 T_y^\dagger=\rho_2$ satisfying
\begin{equation}
    W_x\rho_2=-\alpha_x\rho_2
\end{equation}
Since $\alpha_x\mathrm{Tr}(\rho_1\rho_2)=\mathrm{Tr}(W_x\rho_1\rho_2)=\mathrm{Tr}(\rho_1\rho_2W_x)=-\alpha_x\mathrm{Tr}(\rho_1\rho_2)$, this implies that $\rho_1$ and $\rho_2$ must be orthogonal. Like in the discussion with the toric code pure ground states, there are four density matrices 
\begin{equation}\label{densitymat}
    \{\rho_i\}=\{\rho_1,T_y\rho_1 T_y^\dagger, T_x\rho_1 T_x^\dagger, T_xT_y\rho T_y^\dagger T_x^\dagger\}
\end{equation}
that are orthogonal and locally indistinguishable:
\begin{equation}\label{indmixed}
    \mathrm{Tr}(O\rho_i O^\dagger\rho_j)=c_{O}\delta_{ij}. 
\end{equation}

On the other hand, in the trivial phase where both symmetries are weak, since $W_x\rho_2 W_x^\dagger=\rho_2$, $\rho_1$ and $\rho_2$ need not be orthogonal. In fact, the observation that the weak symmetries are not SSB immediately implies that the four density matrices in (\ref{densitymat}) are locally distinguishable; there is no code property.

The local indistinguishability of the density matrices $\{\rho_i\}$ is directly related to the fact that we can view $\{\rho_i\}$ as pure states under the Choi-Jamiołkowski map, and these pure states satisfy (\ref{indpure}). 

\section{Decohered toric code}
\label{sec:decohered}
In this section we revisit the problem of decohered toric code \cite{fan2024diagnostics, bao2023mixedstate, lee2, Zou_2023, sang2024Markov} to gain more intuition about our definitions of SSB of strong/weak 1-form symmetries, Rényi-2 CMI and SW-SSB. The order and disorder parameters of the Wilson and 't Hooft symmetries have been separately computed in the literature, but we will present a review of the calculation below to be self-contained. 

The initial pure-state density matrix is $
\rho_0=|\psi_0\rangle \langle \psi_0 |$, where $|\psi_0\rangle $ is the fixed-point toric code ground state (\ref{puretc}). The quantum channel we consider is
\begin{equation}
   \mathcal{E}[\rho]=\bigotimes_e\mathcal{E}_e[\rho],\quad  \mathcal{E}_e[\rho]= (1-p)\rho+pZ_e\rho Z_e. 
    \label{Eq: Z dephasing}
\end{equation}

To proceed, we choose a convenient representation of $|\psi_0\rangle$ on a sphere: 
\begin{equation}
|\psi_0\rangle\propto\prod_v\left(\frac{1+A_v}{2}\right)|\uparrow\rangle,
\end{equation}
where $|\uparrow\rangle$ is the state with all spins in the $+1$ eigenstate of $Z$. 
$|\psi_0\rangle$ can be written as an equal superposition of all closed loop states $|g\rangle$, where $|g\rangle$ is a product state with spins in the $-1$ eigenstate of $Z_e$ only along the closed loops $g$:
\begin{equation}
    |\psi_0\rangle=\frac{1}{2^{(N_v-1)/2}}\sum_{g}|g\rangle.
\end{equation}
$\rho_0$ in the doubled Hilbert space is thus
\begin{equation}
|\rho_0\rrangle=|\psi_0\rangle|\psi_0^*\rangle\propto\sum_{g_u,g_l}|g_u,g_l\rrangle,
\end{equation}
where $g_u$ and $g_l$ are independent loop configurations in the upper and lower Hilbert spaces. 
The channel $\E[\rho_0]$ sends the Choi state to
\begin{align}
    |\rho\rrangle&=e^{\mu\sum Z_uZ_e}|\rho_0\rrangle\propto\sum_{g_u,g_l}e^{-2\mu|g_u+g_l|}|g_u,g_l\rangle,
\end{align}
where $\mu=\tanh^{-1}\left(\frac{p}{1-p}\right)$. Here, $|g_u+g_l|$ is the difference between loops $g_u$ and $g_l$, i.e. the loop given by $\prod_{l'\in g_l}X_{l'}\prod_{l\in g_u}X_l|\uparrow\rangle$. 
Relabeling $g=g_u$ and $h=g_u+g_l$, 
\begin{equation}
    |\rho\rrangle\propto
    |\psi_0\rangle \sum_{h}e^{-2\mu |h|}|h\rangle. 
\label{eq:Choi_Kim}
\end{equation}
 Note that the above change of basis is identical to the transformation we did in Sec.~\ref{schoitemp}. 
\eqref{eq:Choi_Kim} is nothing but a pure state toric code trivially stacked with a wave-function deformed toric code, and the latter was examined in \cite{Deformed}. It was shown there that closed $X$ loops always decays as perimeter law for any $\mu$, which in our setup translates into the fact that the order parameter $O_1$ of the weak Wilson symmetry always scales as perimeter law regardless of the value of $p$. Alternatively one can obtain the same result using the fact that $O_1$ is related to the free energy of an Ising domain wall in the dual model, see for example \cite{bao2023mixedstate}. This seems to indicate that the weak Wilson symmetry is always spontaneously broken.

However, one should be careful and further check the disorder parameter $D_2,$ which reads 
\be
\begin{split}
& \frac{\tr (\rho \prod_{e\in {\gamma}} Z_e \rho \prod_{e'\in {\gamma}} Z_{e'}) }{\tr \rho^2} = \frac{\llangle \rho | \prod_{e\in {\gamma}} Z_e^u  Z_e^l |\rho \rrangle }{ \llangle \rho | \rho \rrangle}\\
= & \frac{\sum_h \langle h| \prod_{e\in \gamma} Z_e e^{-4\mu |h|}|h\rangle}{ \sum_h \langle h| e^{-4\mu |h|} | h\rangle}\\
\end{split}.
\ee
Here $\gamma$ is an open string on the lattice. We can interpret the above as dual to the spin-spin correlator in the 2+0d classical Ising model on a square lattice, where the two spin operators live on the endpoints of $\gamma$, which is long-ranged above the critical temperature. If $\mu\to\infty$ (temperature goes to zero), the disorder parameter is 1.  Therefore, above a critical error rate $p>p_c^{(2)}$, the disorder parameter is $\mathcal{O}(1)$, indicating the weak Wilson symmetry is still present. 

Next we turn to the strong Wilson symmetry. Its order parameter $O_2$ is straightforward to evaluate using the  fact that closed $X$-dual loops commute with the channel:
\be
\begin{split}
&  \frac{\tr (\rho \prod_{e\in \tilde{\gamma}} X_e \mathcal{E}[\rho_0] \prod_{e'\in \tilde{\gamma}} X_{e'}) }{\tr \rho^2}\\
= & \frac{\tr (\rho  \mathcal{E}[\prod_{e\in \tilde{\gamma}} X_e \rho_0 \prod_{e'\in \tilde{\gamma}} X_{e'} ])}{\tr \rho^2} = \frac{\tr (\rho  \mathcal{E}[\rho_0])}{\tr \rho^2}=1.\\
\end{split}
\ee
Here $\tilde{\gamma}$ is a closed loop on the dual lattice. 
Since the above holds for all values of $\mu$ or error rate $p$, the strong  Wilson symmetry is broken for any $p$. 

It is also straightforward to compute $D_1$. For any $p<\frac{1}{2}$, which corresponds to (strictly) finite time local Lindbladian evolution, $D_1$ is always zero because it commutes with the channel.

More generally, finite time local Lindbladian evolution is known to obey Lieb-Robinson bounds \cite{Poulin_2010} (see also \cite{Anthony_Chen_2023}): 
\begin{equation}\label{LRbound}
    \|\left[e^{\mathcal{L}t}[O_i],O_j\right]\|\leq e^{\mu (v_{LR}t-|i-j|)},
\end{equation}
where $O_{i(j)}$ is an operator fully supported on regions $i$ ($j$), $v_{LR}$ and $\mu$ are constants that depend on locality and norms of local terms in the Lindbladian $\mathcal{L}$ and $|i-j|$ is the distance between $i$ and $j$. Here $||O||$ is the operator norm defined as $||O||=\sup_{\|\psi\|=1} ||O \psi ||.$ This means that local Lindbladian evolution can never cause a diverging correlation length for \emph{linear} correlation functions, since $\mathrm{Tr}(\rho O)\leq\|O\|$ for an apropriately normalized $\rho$. Therefore, we do not expect linear order and disorder parameters to change in scaling. 

Nevertheless, we can still use local Lindbladian evolution to go from the toric code state to a SW-SSB state because the Lieb-Robinson bound does \emph{not} constrain the R\'enyi-2 correlators. In particular, the correlation length for R\'enyi-2 connected correlation functions (i.e. Markov length) can diverge even along a path generated by local Lindbladian evolution. For example, $D_2$ demonstrates a diverging correlation length even after finite time local Lindbladian evolution, as shown above.

To summarize for decohered toric code, although $O_1$ decays with a perimeter law, which is unusual for pure states with a long-ranged disorder parameter, the above state for $p>p_c^{(2)}$ is equivalent to the SW-SSB fixed point according to our equivalence relation. Therefore, it describes the SW-SSB phase for $p>p_c^{(2)}$. The superscript $(2)$ indicates Rényi-2 error threshold, which is different from the von Neumann error threshold $p_c^{(1)}$ as explained in Sec.~\ref{sdiscussion} in the main text.

\subsection{R\'enyi-2 Markov length}\label{si2tc}
In this appendix, we compute the R\'enyi-2 Markov length explicitly for toric code decohered in both $Z$ and $X$ channels, and show that in this example the R\'enyi-2 Markov length diverges at two points. These points separate the ST-SSB phase from the SW-SSB phase and the SW-SSB phase from the WS phase. Our results are analogous to those of Ref.~\cite{sang2024Markov}, except with random bond Ising model replaced by the usual Ising model.

The density matrix for the pure toric code state on a sphere is
\begin{equation}
    |tc\rangle\langle tc|=4\prod_p\left(\frac{1+B_p}{2}\right)\prod_v\left(\frac{1+A_v}{2}\right)
\end{equation}
under $Z$ decoherence with amplitude $p_z$ and $X$ decoherence with amplitude $p_x$, we get
\begin{align}
    \begin{split}  \label{decrho}  
\rho&=\frac{4}{2^{N_v+N_p}}\left(\sum_{\sigma}\prod_pB_p^{\sigma_p}(1-2p_p)^{N_{\sigma}}\right)\\
    &\times\left(\sum_{\sigma'}\prod_vA_v^{\sigma_v'}(1-2p_v)^{N_{\sigma'}}\right)
\end{split}
\end{align}
where $\sigma$ and $\sigma'$ are vectors with $N_p$ and $N_v$ elements $\sigma_p=0,1$, $\sigma_v'=0,1$ respectively, and we sum over all such vectors. Note that the Pauli $X$ and Pauli $Z$ parts of the density matrix completely decouple.

If we trace over a region $C$, if $\bar{C}$ is simply connected, then we get
\begin{align}
\mathrm{Tr}_C(\rho)&=\frac{1}{2^{N_{v,\bar{C}}+N_{p,\bar{C}}}}\left(\sum_{\sigma\in {\bar{C}}}\prod_pB_p^{\sigma_p}(1-2p_p)^{N_{\sigma}}\right)\nonumber\\
&\times\left(\sum_{\sigma'\in\bar{C}}\prod_vA_v^{\sigma_v'}(1-2p_v)^{N_{\sigma'}}\right),
\end{align}
where $\sigma\in\bar{C}$ and $\sigma'\in\bar{C}$ are restricted to be nonzero for plaquettes and vertices fully supported in $\bar{C}=A\cup B$. 
It follows that
\begin{align}
\begin{split}
    &\mathrm{Tr}(\mathrm{Tr}_C(\rho)^2)\\
    &=\frac{1}{2^{N_{v,\bar{C}}+N_{b,\bar{C}}}}\left(\sum_{\sigma\in\bar{C}}(1-2p_p)^{2N_{\sigma}}\right)\left(\sum_{\sigma'\in\bar{C}}(1-2p_v)^{2N_{\sigma'}}\right)\\
    &=\frac{1}{2^{N_{v,\bar{C}}+N_{b,\bar{C}}}}\left(\sum_{\sigma\in\bar{C}}e^{-4\beta_pN_{\sigma}}\right)\left(\sum_{\sigma'\in\bar{C}}e^{-4\beta_vN_{\sigma'}}\right)\\
    &\propto Z_{\bar{C}}(\beta_p)Z_{\bar{C}}(\beta_v)
\end{split}
\end{align}
where $\tanh(\beta_p)=\frac{p_p}{1-p_p}$ and $\tanh(\beta_v)=\frac{p_v}{1-p_v}$. Here we used the notation $Z_{\bar{C}}(\beta_p)$ to denote the 2D classical Ising partition function $\sum_{\sigma'\in\bar{C}}e^{-4\beta_vN_{\sigma'}}$ for region $\bar{C}$. 

For a region that is not simply connected, we get two contributions:
\begin{align}
\begin{split}
    &\mathrm{Tr}(\mathrm{Tr}_{AC}(\rho)^2)\\
    &\propto (Z_{B,e}(\beta_p)+Z_{B,o}(\beta_p))(Z_{B,e}(\beta_v)+Z_{B,o}(\beta_v))
\end{split}
\end{align}
where $Z_{B,e}(\beta_p)$ is a sum over configurations with an even number of domain walls between the inner and outer edges of $B$ and $Z_{B,o}(\beta_p)$ is a sum over configurations with an odd number of domain walls between the inner and outer edges of $B$. Plugging the above results into (\ref{Rényi2cmi}) gives
\begin{align}
\begin{split}
&I_2(A:C|B)\\
&=-\log\frac{Z_{A\cup B}(\beta_p)(Z_{B\cup C,e}(\beta_p)+Z_{B\cup C,o}(\beta_p))}{(Z_{B,e}(\beta_p)+Z_{B,o}(\beta_p))Z_{A\cup B\cup C}(\beta_p)}\\
&-\log \frac{Z_{A\cup B}(\beta_v)(Z_{B\cup C,e}(\beta_v)+Z_{B\cup C,o}(\beta_v))}{(Z_{B,e}(\beta_v)+Z_{B,o}(\beta_v))Z_{A\cup B\cup C}(\beta_v)}\\
&=F(4r,\beta_p)-F(2r,\beta_p)+F(4r,\beta_v)-F(2r,\beta_v)
\end{split}
\end{align}
where $F(x,\beta)$ is the free energy cost of introducing a point defect in the 2D classical Ising model of linear size $x$ at temperature $\beta$. In the above, we chose $A$ to be very small compared to $r$, so we can treat it as a point. We see that $I_2(A:C|B)$ is the sum of two contributions, which depend on $\beta_p$ and $\beta_v$ respectively. Because we can tune $\beta_p$ and $\beta_v$ independently, we get two transitions, one when $\beta_p$ reaches the 2D classical Ising critical temperature and the other when $\beta_v$ reaches the 2D classical Ising critical temperature. At both of these points, $I_2(A:C|B)$ decays with a power law rather than an exponential, so the R\'enyi-2 Markov length diverges.

\section{Sequential circuit for $|\rho(\beta)\rrangle$}
\label{ssequcirc}
In this appendix, we will show how to derive an effective unitary $U(\beta)$ satisfying
\begin{equation}
    |\rho(\beta)\rangle=U(\beta)|\uparrow\rangle
\end{equation}
where 
\begin{equation}
    |\rho(\beta)\rangle=\frac{1}{\cosh^{N_v}(2\beta)}\prod e^{\beta A_v}|\uparrow\rangle
    \end{equation}
on a system with open boundary conditions. The trick we use is similar to that used in Ref.~\onlinecite{jones2021}: we pull $Z_e$ operators out from $|\uparrow\rangle$ to turn $\prod e^{\beta_AA_v}$ into a sequential circuit on this particular state. This sequential circuit is exact for open boundary conditions in at least one direction. Closing up the boundary seems a bit tricky though, at least if we want to stay exact. We will show that this circuit maps local operators to exponentially localized operators. We can use this unitary operator to obtain a quasilocal parent Hamiltonian that has a constant gap for all (finite) $\beta$ because it is just $H_B$ conjugated by a unitary operator. 

\begin{figure}[tb]
   \centering
   \includegraphics[width=.8\columnwidth]{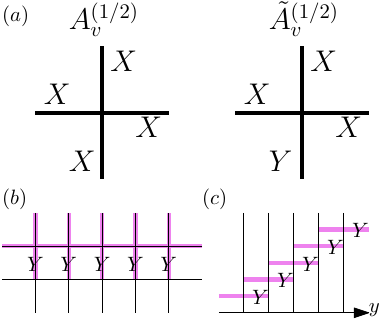} 
   \caption{Unitary circuit relating $|\rho_B\rrangle $ to $|\rho_{\beta}\rrangle$ when we take one of the directions (say, $y$) to have open boundary conditions. (a) We can write $e^{\beta A_v}|\uparrow_{m}\rangle$ as $e^{i\tilde{\beta}\tilde{A}_{v,m}}|\uparrow_{m}\rangle$ where $\tilde{A}_{v,m}$ has a single $Y_{e,m}$ operator. (b) We can transform all the $A_{v,m}$ terms in a single row ($x$ direction) simultaneously. (c) Side view of the circuit. The circuit is a sequential circuit in the $y$ direction because the gates in different $y$ coordinates do not commute. The $Y_{e,m}$ operator of the $\tilde{A}_{v,m}$ one row below overlaps with a $X_{e,m}$ operator of the $\tilde{A}_{v,m}$ one row above. The naive construction illustrated here allows periodic boundary conditions in the $x$ direction but not in the $y$ direction; closing up the circuit in the $y$ direction would require some modifications.}
   \label{fig:circuitAv}
   \end{figure}
   
We can turn the non-unitary operator $e^{\beta A_v}$ into a unitary operator as follows. We use
\begin{align}
    \begin{split}
        e^{\beta A_v}|\uparrow\rangle&=(\cosh \beta+\sinh\beta A_v)|\uparrow\rangle\\
        &=(\cosh\beta Z_e+\sinh\beta A_vZ_e)|\uparrow\rangle\\
        &=(\cosh\beta-i\sinh\beta \tilde{A}_v)|\uparrow\rangle
    \end{split}
\end{align}
where $Z_e$ overlaps with $A_v$, so $\tilde{A}_v$ has a single $Y_e$ operator. This operator is illustrated in Fig.~\ref{fig:circuitAv}.(a). Since $\tilde{A}_v$ is Hermitian, we can write
\begin{equation}
    e^{\beta A_v}|\uparrow\rangle=e^{i\tilde{\beta}\tilde{A}_v}|\uparrow\rangle
\end{equation}
where
\begin{equation}\label{betatilde}
    \tan(\tilde{\beta})=-\tanh(\beta)
\end{equation}
Therefore, by pulling out $Z_e$ operators from $|\uparrow\rangle$, we can turn a single non-unitary gate $e^{\beta A_v}$ into a unitary operator $e^{i\tilde{\beta}\tilde{A}_v}$, which has the same action on the state $|\uparrow\rangle$ up to a proportionality constant. Now we need to perform this transformation on every site. Note that we can do this transformation in one direction (say, the $x$ direction) all at once (see Fig.~\ref{fig:circuitAv}.(b)), but then we need to do this transformation in the other direction (say, the $y$ direction) sequentially (see Fig.~\ref{fig:circuitAv}.(c)), because the gates after this transformation no longer commute. Therefore, we end up with a circuit that is sequential in one direction. We can have periodic boundary conditions in the $x$ direction, but modifying the circuit to have periodic boundary conditions in the $y$ direction seems trickier (at least for an exact circuit solution). 

Putting this together, the above discussion gives an exact (for open boundary conditions in the $y$ direction) sequential circuit $U(\beta)$ given by
\begin{align}
\begin{split}
&U(\beta)\\
&=\prod_{v_x,v_y=L-2} e^{i\tilde{\beta}\tilde{A}_v}\dots\prod_{v_x,v_y=1} e^{i\tilde{\beta}\tilde{A}_v}\prod_{v_x,v_y=0} e^{i\tilde{\beta}\tilde{A}_v}
\end{split}
\end{align}

Because the circuit is sequential in the $y$ direction, it gives a correlation length in the $y$ direction. In other words, $U(\beta)O_eU^\dagger(\beta)$ generically maps a local operator $O_e$ supported on site $e$ to an operator with strict locality in the $x$ direction but an exponential tail in the $+y$ direction (note that there is no tail in the $-y$ direction). 

Next, we must show that $ U(\beta)Z_eU^\dagger(\beta)=\tilde{Z}_e$ is exponentially localized. Since $\llangle\rho_B| Z_eZ_{e'}|\rho_{B}\rrangle=1$, this would also show that $\llangle\rho_{\beta}|\tilde{Z}_e\tilde{Z}_{e'}|\rho_{\beta}\rrangle=1$, i.e. there is an exponentially localized dressed order parameter with expecation value 1. 

We first consider the case where $Z_e$ is supported on a vertical edge, so it overlaps with the vertex terms on $(x,y)$ and $(x,y+1)$. The other case can be handled similarly. We compute
\begin{widetext}
\begin{align}
    \begin{split}    U_{\beta}Z_eU_{\beta}^\dagger&= \left[\prod_{v_x,v_y=0}^{v_y=L-2}e^{i\tilde{\beta}\tilde{A}_v}\right]Z_e\left[\prod_{v_x,v_y=L-2}^{v_y=0}e^{-i\tilde{\beta}\tilde{A}_v}\right]\\
    &= \left[\prod_{v_x=x,v_y=y}^{v_y=L-2}e^{i\tilde{\beta}\tilde{A}_v}\right]Z_e\left[\prod_{v_x=x,v_y=L-2}^{v_y=y}e^{-i\tilde{\beta}\tilde{A}_v}\right]\\
    &=\left[\prod_{v_x=x,v_y=y+1}^{v_y=L-2}e^{i\tilde{\beta}\tilde{A}_v}\right]Z_ee^{-2i\tilde{\beta}\tilde{A}_{x,y}}\left[\prod_{v_x=x,v_y=L-2}^{v_y=y+1}e^{-i\tilde{\beta}\tilde{A}_v}\right]
\end{split}
\end{align}
In the above, we first commuted through the rows labeled by vertices with $v_y<y$ though $Z_e$. We also commuted through the vertices with $v_x\neq x$. Then we pulled $e^{-\tilde{\beta}\tilde{A}_v}$ through $Z_e$ but because these operators do not commute, we have an operator left over. We can simplify the above expression by expanding in terms of cosines and sines:
\begin{align}
    \begin{split}
        U_{\beta}Z_eU_{\beta}^\dagger&=\left[\prod_{v_x=x,v_y=y+1}^{v_y=L-2}e^{i\tilde{\beta}\tilde{A}_v}\right]Z_e\left(\cos(2\tilde{\beta})-i\sin(2\tilde{\beta})\tilde{A}_{x,y}\right)\left[\prod_{v_x=x,v_y=L-2}^{v_y=y+1}e^{-i\tilde{\beta}\tilde{A}_v}\right]\\
&=\left[\prod_{v_x=x,v_y=y+2}^{v_y=L-2}e^{i\tilde{\beta}\tilde{A}_v}\right]Z_ee^{-2i\tilde{\beta}\tilde{A}_{x,y+1}}\left(\cos(2\tilde{\beta})-i\sin(2\tilde{\beta})\tilde{A}_{x,y}\right)\left[\prod_{v_x=x,v_y=L-2}^{v_y=y+2}e^{-i\tilde{\beta}\tilde{A}_v}\right]\\
&=\left[\prod_{v_x=x,v_y=y+2}^{v_y=L-2}e^{i\tilde{\beta}\tilde{A}_v}\right]Z_e\left(c_{\tilde{\beta}}-is_{\tilde{\beta}}\tilde{A}_{x,y+1}\right)\left(c_{\tilde{\beta}}-is_{\tilde{\beta}}\tilde{A}_{x,y}\right)\left[\prod_{v_x=x,v_y=L-2}^{v_y=y+2}e^{-i\tilde{\beta}\tilde{A}_v}\right]\\
&=\left[\prod_{v_x=x,v_y=y+2}^{v_y=L-2}e^{i\tilde{\beta}\tilde{A}_v}\right]Z_e\left(c_{\tilde{\beta}}^2-ic_{\tilde{\beta}}s_{\tilde{\beta}}\tilde{A}_{x,y}-ic_{\tilde{\beta}}s_{\tilde{\beta}}\tilde{A}_{x,y+1}+s_{\tilde{\beta}}^2\tilde{A}_{x,y}\tilde{A}_{x,y+1}\right)\left[\prod_{v_x=x,v_y=L-2}^{v_y=y+2}e^{-i\tilde{\beta}\tilde{A}_v}\right]
\end{split}
\end{align}
where we used the short hand $c_{\tilde{\beta}}=\cos(2\tilde{\beta})$ and $s_{\tilde{\beta}}=\sin(2\tilde{\beta})$. We can now see the exponential tails forming. If $\beta=0$, then $\tilde{\beta}=0$ and $c_{\tilde{\beta}}=1$ while $s_{\tilde{\beta}}=0$. $Z_{e,m}$ remains unchanged. On the other hand, for $\beta\to\infty,\tilde{\beta}\to -\frac{\pi}{4}$ so $c_{\tilde{\beta}}=0$ while $s_{\tilde{\beta}}=-1$ and $Z_e$ becomes delocalized with correlation length $\to\infty$.

The next step is pulling $e^{i\tilde{\beta}\tilde{A}_{x,y+2}}$ through. It commutes with $Z_e$ and $\tilde{A}_{x,y}$ but fails to commute with the other two terms. Therefore, we get in the middle of the two big products
\begin{align}
    \begin{split}    &Z_e\left(c_{\tilde{\beta}}^2-ic_{\tilde{\beta}}s_{\tilde{\beta}}\tilde{A}_{x,y}-ic_{\tilde{\beta}}s_{\tilde{\beta}}\tilde{A}_{x,y+1}e^{-2i\tilde{\beta}\tilde{A}_{x,y+2}}+s_{\tilde{\beta}}^2\tilde{A}_{x,y}\tilde{A}_{x,y+1}e^{-2i\tilde{\beta}\tilde{A}_{x,y+2}}\right)\\
    &=Z_e\left(c_{\tilde{\beta}}^2-ic_{\tilde{\beta}}s_{\tilde{\beta}}\tilde{A}_{x,y}-ic_{\tilde{\beta}}s_{\tilde{\beta}}\tilde{A}_{x,y+1}(c_{\tilde{\beta}}-is_{\tilde{\beta}}\tilde{A}_{x,y+2})+s_{\tilde{\beta}}^2\tilde{A}_{x,y}\tilde{A}_{x,y+1}(c_{\tilde{\beta}}-is_{\tilde{\beta}}\tilde{A}_{x,y+2})\right)
\end{split}
\end{align}

Continuing in this way, we find that 
\begin{equation}
    U_{\beta}Z_eU_{\beta}^\dagger=Z_e\sum \alpha_iO_i
\end{equation}
where $\{O_i\}$ are all connected strings of $A_{v,m}$ operators extending from $v_y=y$ to $v_y=L-2$ with $v_x=x$ and $\alpha_i$ are their coefficients, which are combinations of $c_{\tilde{\beta}}$ and $s_{\tilde{\beta}}$. A string extending from $v_y=y$ to $v_y=y'$ must have a coefficient of at most $c_{\tilde{\beta}}s_{\tilde{\beta}}^{|y'-y-1|}$ or $s_{\tilde{\beta}}^{|y'-y|}$. Taking the limit $L\to\infty$, we find that for an operator $O_e$ supported on the same $x$ coordinate with a different $y$ coordinate $y'$,
\begin{align}
\begin{split}
    \|[\tilde{Z}_e,O_e]\|&\leq |c_{\tilde{\beta}}|\sum_{n=|y'-y-1|}^{\infty}|s_{\tilde{\beta}}|^n+\sum_{n=|y'-y|}^{\infty}|s_{\tilde{\beta}}|^n\\
    &\leq \frac{|c_{\tilde{\beta}}||s_{\tilde{\beta}}|^{|y'-y-1|}}{1-|s_{\tilde{\beta}}|}+\frac{|s_{\tilde{\beta}}|^{|y'-y|}}{1-|s_{\tilde{\beta}}|}\leq \mathrm{const}\cdot e^{-|y'-y|/\xi}
\end{split}
\end{align}
\end{widetext}
where the correlation length $\xi$ in the $y$ direction is given by
\begin{equation}\label{corrlen}
    \xi=-\frac{1}{\log\left[|\sin(2\tilde{\beta})|\right]}
\end{equation}
which diverges as $\beta\to\infty$  ($\tilde{\beta}\to-\frac{\pi}{4})$ and goes to zero as $\beta\to 0$ 
 ($\tilde{\beta}\to 0)$. The spread of $\tilde{Z}_{e}$ where $e$ is a horizontal link can be handled similarly, and has the same correlation length.

 In summary, we find a gapped parent Hamiltonian for the Choi state of $\rho_{\beta}$ that is local, with a finite correlation length, for all finite $\beta$. However in the limit $\beta\to\infty$, where the Choi state has $2\log 2$ TEE rather than just $\log 2$ TEE, the correlation length diverges and the parent Hamiltonian becomes nonlocal

\subsection{Logarithmic time Lindbladian evolution}\label{slogtime}

 One nice aspect of the above derivation of $\xi$ as a function of $\tilde{\beta}$ is that it allows us to show that there is a logarithmic time Lindbladian evolution to get from the trivial state to the SW-SSB state. First, we can use the amplitude damping channel described in Appendix~\ref{app:channel} to get from the infinite temperature state to the product state $\rho=\prod_e\left(\frac{1+Z_e}{2}\right)$. Then to get from the product state to (\ref{SW-SSBz2}), we apply the channel
 \begin{equation}
     \rho\to(1-p)\rho+pA_v\rho A_v
 \end{equation}
 on every vertex $v$. At $p=\frac{1}{2}$, the resulting density matrix is precisely (\ref{SW-SSBz2}). The Lindbladian evolution corresponding to this channel is
 \begin{equation}
     \mathcal{L}[\rho]=\frac{1}{2}\sum_v(A_v\rho A_v-\rho)
 \end{equation}
 with $t_p=-\log(1-2p)$. Clearly, $p=\frac{1}{2}$ requires infinite $t_p$. However, we can instead take $t_p=\log L$. This gives 
 \begin{equation}
     p=\frac{1}{2}(1-e^{-t})=\frac{1}{2}\left(1-\frac{1}{L}\right)
 \end{equation}
Using $(1-2p)=e^{-2\beta}$ gives
 \begin{equation}
     \beta=-\frac{1}{2}\log\frac{1}{L}
 \end{equation}
 Then using (\ref{betatilde}) and (\ref{corrlen}), we obtain in an expansion in large $L$,
 \begin{equation}
 \xi=\frac{1}{2}L^2
 \end{equation}

Therefore, $\xi$ is no longer finite even if we choose $t_p$ to scale only as $\log L$! Since $\xi$ can sufficiently diverge even if $t_p=\log L$, there is no obstruction to using Lindbladian evolution up to time $\sim \log L$ to get to the SW-SSB phase from the infinite temperature state. A similar calculation with $t_p=L$ would give $\xi\sim\frac{1}{2}e^{2L}$.

\section{Proof of exact 't Hooft symmetry}\label{sweakthooft}

In this appendix, we will prove that the ensemble of ground states for the disordered Hamiltonian in Eq. \ref{hamrandfield} has an exact weak 't Hooft 1-form symmetry, i.e.
\begin{equation}
    \prod_{i\in \ell}X_e\rho\prod_{i\in \ell}X_e=\rho
\label{eq:field_disorder_2nd}
\end{equation}
for any closed loop $\ell$ on the dual lattice. In fact, the LHS can be regarded as the ensemble of ground states of the following disordered Hamiltonian
\begin{equation}
    H(\{h_e^X\})=\prod_{i\in \ell}X_e H(\{h_e\})\prod_{i\in \ell}X_e,
\end{equation}
with a new set of random couplings 
\be
h^X_e=\begin{cases}  -h_e, & \text{if}\ e \in\ell,\\
 h_e, & \text{otherwise.}
\end{cases}
\ee
Since each $h_e$ are independently drawn from a bimodal distribution $h_e=\pm h$ or Gaussian distribution with zero mean, $-h_e$ follows the same probability distribution as $h_e$. Therefore, the Hamiltonian $H(\{h_e^X\})$ describes the same disordered system, and the two density matrices in \eqnref{eq:field_disorder_2nd} are hence equal to each other.

\section{Quasi-adiabatic continuation and derivation of $U(\{h_e\})$}\label{squasi}

In this section we derive the disordered density matrix using quasi-adiabatic continuation for bimodal disorder.

\subsection{ST-SSB phase}
We can approximate the ground state using quasi-adiabatic continuation. We write the ground state as 
\begin{equation}
    |\psi(\{h_e\})\rangle=U|\psi_0\rangle
\end{equation}
where $U=\mathcal{P}e^{i\int_0^1dsD(s)}$ and
\begin{equation}\label{quasiadiabatic}
    D(s)=-i\int dt F(\Delta Et)e^{i H(s)t}\partial_sH(s)e^{-i H(s)t}
\end{equation}
where $H(s)=H_0+s\sum_eh_eZ_e$ so $\partial_sH(s)=\sum_eh_eZ_e$. We assume the presence of a gap $\Delta E$ for all $s\in[0,1]$ for the TO phase. $F(\Delta Et)$ is an antisymmetric function, so that $D(s)$ is Hermitian. Furthermore, it must be maximally quickly decaying given that its Fourier transform $\hat{F}(\omega)=-\frac{1}{\omega}$ for $|\omega|\geq 1$. It can be shown that $F(\Delta Et)$ can be chosen to be (almost) exponentially decaying for $\Delta Et>1$. This guarantees that $D(s)$ is local. \footnote{Specifically, it is local because we can split the integral over $-\frac{1}{\Delta E}\leq t\leq \frac{1}{\Delta E}$ and $\frac{1}{\Delta E}\leq |t|\leq\infty$. Then the first integral is bounded due to Lieb-Robinson bounds and the second integral is bounded due to decay of $F(\Delta Et)$. }

We can commute the $B_p$ operators through $\partial_sH(s)$ so we can replace $H(s)$ by $-\lambda_A\sum A_v+s\sum_eh_eZ_e$. Clearly, $D(s)$ commutes with $\{B_p\}$, so $U$ commutes with $\{B_p\}$.
The density matrix is given by
\begin{align}
\begin{split}
    \rho&\propto\sum_{\{h_e\}}U(\{h_e\})|\psi_0\rangle\langle\psi_0|U^\dagger(\{h_e\})\\
    &\propto\sum_{\{h_e\}}\left(\frac{1+B_p}{2}\right)\left(\frac{1+U(\{h_e\})A_vU^\dagger(\{h_e\})}{2}\right)\\
    &\propto\left(\frac{1+B_p}{2}\right)\sum_{\sigma}U(\{h_e\})\left(\prod_vA_v^{\sigma_v}\right)U^\dagger(\{h_e\})
\end{split}
\end{align}

As long as the gap remains open, $U(\{h_e\})$ is a locality-preserving unitary, so $\mathrm{Tr}(\prod X\rho)$ decays with the length of the loop (i.e. perimeter law). $\rho$ is a sum of states where in every state, $\prod X$ decays with the length of the loop. However, just because every state is in the Toric code phase does not mean that $\rho$ is in the Toric code phase; it can also be in the SW-SSB phase. To be more rigorous, we can work in the limit where the perturbation $\sum_eh_eZ_e$ is small.

First, we write $e^{iH(s)t}$ in (\ref{quasiadiabatic}) as
\begin{equation}
    e^{iH(s)t}=U_A(s,t)U_Z(s,t)
\end{equation}
where
\begin{align}
\begin{split}
    U_Z(s,t)&=e^{is\sum_eh_eZ_et}\\ U_A(s,t)&=\mathcal{T}e^{-i\int_0^tdt'\tilde{H}_A(t')}
\end{split}
\end{align}
and
\begin{align}
\begin{split}
    \tilde{H}_A(s,t')&=U_Z(s,t')H_AU_Z^\dagger(s,t')\\
    &=-\lambda_A\sum_vA_ve^{-2ist'\sum_{l\in v}h_eZ_e}
\end{split}
\end{align}
where the sum is over links that have an endpoint on the vertex $v$. The above can be obtained by trotterizing the time evolution to alternate between evolving with $s\sum h_eZ_e$ and $-\lambda_A\sum A_v$, and then rearranging terms appropriately. Note that flipping $h_e\to-h_e$ does not only take $\partial_sH_s\to-\partial_sH_s$ but also modifies $\tilde{H}_A(s,t')$. Therefore, it does not simply take $D(s)\to -D(s)$ and $U\to U^\dagger$. 

To obtain $D(s)$ to lowest order in $ht\leq \frac{h}{4\lambda_A}$, we can just write $\tilde{H}_A(s,t')\sim -\lambda_A\sum_vA_v+\mathcal{O}\left(\frac{h}{\lambda_A}\right)$. After commuting $U_Z(s,t)$ through $\partial_sH_s$, we get
\begin{widetext}
\begin{align}
\begin{split}
    D(s)&\sim-i\int_{-\infty}^{\infty}dtF(\Delta Et)e^{-i\lambda_A\sum A_vt}\left(\sum h_eZ_e\right)e^{i\lambda_A\sum A_vt}\\
    &=-i\int_{-\infty}^{\infty}\frac{d\tilde{t}}{\Delta E}F(\tilde{t})\sum h_eZ_ee^{\frac{2\lambda_Ai}{\Delta E}(A_v+A_{v'})\tilde{t}}
\end{split}
\end{align}
where $v$ and $v'$ are the endpoints of the link $l$ and we changed the integration variable to $\tilde{t}$ rather than $t$. Simplifying further, we get
\begin{align}
\begin{split}
D(s)&=-\frac{i}{\Delta E}\sum h_eZ_e\int_{-\infty}^{\infty}d\tilde{t}F(\tilde{t})i\cos\left(\frac{2\lambda_A}{\Delta E}\right)\sin\left(\frac{2\lambda_A}{\Delta E}\right)(A_v+A_{v'})\\
&=-\frac{i}{\Delta E}\sum h_eZ_e(A_v+A_{v'})\int_{-\infty}^{\infty}d\tilde{t}F(\tilde{t})\frac{i}{2}\sin\left(\frac{4\lambda_At}{\Delta E}\right)
\end{split}
\end{align}
where we dropped terms in the expansion of the exponential that are even in $\tilde{t}$ because $F(\tilde{t})$ is an odd function of $\tilde{t}$. Using $\Delta E=4\lambda_A$ and writing the sine in terms of exponentials, we get
\begin{equation}
D(s)=-\frac{i}{4\lambda_A}\sum h_eZ_e(A_v+A_{v'})\int_{-\infty}^{\infty}d\tilde{t}F(\tilde{t})\frac{1}{4}\left(e^{i\tilde{t}}-e^{-i\tilde{t}}\right)
\end{equation}
\end{widetext}

Finally, we use the Fourier transform property $\tilde{F}(\omega)=-\frac{1}{\omega}$ with $\omega=1$. This gives

\begin{equation}\label{Deq}
D(s)=\frac{i}{8\lambda_A}\sum h_eZ_e(A_v+A_{v'})
\end{equation}

Since there are two terms for each $Z_e$ and $A_v,A_{v'}$ get absorbed into $|\psi_0\rangle$, we see that this reproduces the result from perturbation theory! Specifically, expanding $U|\psi_0\rangle=e^{iD}|\psi_0\rangle\sim 1+iD|\psi_0\rangle$ at first order gives precisely the result from first order perturbation theory on the wavefunction. This form of $D(s)$ is reasonable -- it is the simplest $D(s)$ that reproduces perturbation theory that is local, Hermitian, and symmetric with respect to symmetries of the square lattice.

Notice that if instead we had Gaussian disorder, then $U$ would only be modified locally, i.e. it would deviate from $e^{iD}$ with D given by (\ref{Deq}) in small local regions, where we cannot expand in $\frac{h}{\lambda_A}$. In this case, we can write $U_{\mathrm{gauss}}=\prod U_i U$ where $U_i$ is localized near a site $i$ and are sparse, occurring on a fraction $\sim 1-\mathrm{Erf}(\lambda/\Delta_h)$ of the sites. The crucial assumption here is that the $\{U_i\}$ are local and sparse, so we can approximate $\prod_iU_i$ by a product of disjoint unitaries.

Now we evaluate
\begin{widetext}
\begin{align}
\begin{split}
    \rho&\propto\prod\left(\frac{1+B_p}{2}\right)\sum_{\sigma}\prod A_v^{\sigma_v}\left(\prod A_v^{\sigma_v}U(\{h_e\})\prod A_v^{\sigma_v}\right)U^\dagger(\{h_e\})\\
    &\propto\prod\left(\frac{1+B_p}{2}\right)\sum_{A}\left(\prod_{l\in\partial A} X_l\right)\prod_{l\in\partial A}e^{-i\frac{ih_eZ_e(A_{v}+A_{v'})}{4\lambda_A}}
\end{split}
\end{align}
where we used $e^{t(A+B)}=e^{tA}e^{tB}e^{-(t^2/2)[A,B]}\cdots$ and dropped terms higher order in $t=-\frac{h}{2\lambda}$. For Gaussian disorder, there would be an additional conjugation by $\prod_iU_i$. However, because the $\{U_i\}$ are local and disjoint, we can commute most of them through $\left(\prod_{l\in\partial A} X_l\right)\prod_{l\in\partial A}e^{-i\frac{ih_eZ_e(A_{v}+A_{v'})}{4\lambda_A}}$. Only the $\{U_i\}$ with support on the loop $\partial A$ remain, and they will modify the perimeter law slightly. 

If we write
\begin{equation}
    \prod_{l\in\partial A}e^{-i\frac{ih_eZ_e(A_{v}+A_{v'})}{4\lambda_A}}=\prod_{l\in\partial A}\left(\cos\left(\frac{h_e}{4\lambda_A}\right)-i\sin\left(\frac{h_e}{4\lambda_A}\right)iZ_eA_{v}\right)\left(\cos\left(\frac{h_e}{4\lambda_A}\right)-i\sin\left(\frac{h_e}{4\lambda_A}\right)iZ_eA_{v'}\right)
\end{equation}
 Then all of the terms with at least one sine will cancel when we sum over $\{h_e\}$. Therefore, all that is left is 
\begin{equation}
    \sum_{\{h_e\}}\prod_{l\in\partial A}e^{i\frac{ih_eZ_e(A_{v}+A_{v'})}{4\lambda_A}}\sim\prod_{l\in\partial A}\left[\cos^2\left(\frac{h_e}{4\lambda_A}\right)-\sin^2\left(\frac{h_e}{4\lambda_A}\right)A_vA_{v'}\right]
\end{equation}
Any term with a $\sin^2\left(\frac{h_e}{4\lambda_A}\right)A_vA_{v'}$ will be much smaller than the term with only $\cos^2\left(\frac{h_e}{4\lambda_A}\right)$. Therefore, we can keep only the term
$\cos^{2|\partial A|}\left(\frac{h}{4\lambda_A}\right)$. For small $\frac{h}{4\lambda}$, this gives
\begin{equation}
    \rho\propto\prod\left(\frac{1+B_p}{2}\right)\sum_{\sigma}\left(\prod_vA_v^{\sigma}\right)\left(1-\frac{h^2}{16\lambda^2}\right)^{2N_{\sigma}}
\end{equation}
\end{widetext}
which matches with the decohered toric code (\ref{decrho}) at $p_p=0$ and 
\begin{equation}
    1-2p_v=\left(1-\frac{h^2}{16\lambda_A^2}\right)^2\sim1-\frac{h^2}{8\lambda_A^2}
\end{equation}

\subsection{WS phase}

Note that we could have also approached this problem using quasi-adiabatic continuation. In this case, we find that 
\begin{equation}
    D=-\frac{i}{8h}\lambda_A\sum A_v\frac{1}{8}\left(\sum s_lZ_e+\sum s_ls_{l'}s_{l''}Z_eZ_{l'}Z_{l''}\right)
\end{equation}
where the sum is over links with an endpoint on the vertex $v$ and $s_l=\mathrm{sign}(h_e)$. Then we find
\begin{equation}
    \rho\propto\sum e^{iD}\prod_ls_lZ_ee^{-iD}
\end{equation}
After summing over $\{s_l\}$ removes all the terms except those for with a term in $D$ cancels with $\prod_ls_lZ_e$. To leading order, the only terms for which such a cancellation occurs are single $Z_e$ operators and $Z_eZ_{l'}Z_{l''}$ operators around a vertex. This gives, to leading order in $\frac{\lambda_A}{4h}$,
\begin{equation}
\rho\propto 1+\frac{\lambda_A}{4h}\sum_vA_v
\end{equation}
which matches with (\ref{rhorp}).

\section{Scaling of $O_{1,z}$ for Gaussian disorder}\label{so1zgauss}

Here we consider the scaling of the quantity $\mathrm{Tr}(\prod_{e\in\ell}Z_e\rho)$ 
with the size of the loop $\gamma$. The scaling of $\mathrm{Tr}(W(\gamma)\rho)$ should match with that of $\mathrm{Tr}(W(\gamma)\rho^2)/\mathrm{Tr}(\rho^2)$ because $\mathrm{\rho}$ and $\mathrm{\rho}^2$ only differ by squaring the eigenvalues of $\rho$ and renormalizing; the eigenstates are the same.\footnote{Note that this argument only applies for linear correlators. Nonlinear correlators like R\'enyi-2 correlators can demonstrate a change in scaling after taking $\rho\to\rho^2$.}

In the limit that $\lambda_B\to 0$, we have
\begin{equation}
    \mathrm{Tr}(\rho B_p)=0
\end{equation}
for all $B_p$. This comes from the fact that the distribution of $\{h_e\}$ have zero mean. Now we turn a small $B_p$ term, so that $\frac{\lambda_B}{\Delta_h}\ll 1$. This gives a small probability for the $B_p$ term to overpower the $Z_e$ terms for a given plaquette. Specifically, for a particular disorder realization, if the $Z_e$ terms in a plaquette favor $B_p=-1$ but there is a link with $h_e<\lambda_B$, then it would be energetically favorable for $B_p=-1$ to flip to $B_p=+1$ after the addition of the $-\lambda\sum B_p$ term to the Hamiltonian. The probability for a $B_p$ term to overpower the $Z_e$ terms is $\sim\frac{\sqrt{2}\lambda_B}{\sqrt{\pi}\Delta_h}$ as can be obtained from integrating over the Gaussian distribution $\frac{1}{\sqrt{2\pi}\Delta_h}\int_{-\lambda_B}^{\lambda_B}dh\ e^{-h^2/2\Delta_h^2}$ and expanding at small $\lambda_B/\Delta_h$. 

$W(\gamma)$ can be written as $W(\gamma)=\prod_{p\in \gamma}B_p$ (we now treat $\gamma$ as a disk). In a particular disorder realization, flips of $B_p$ at the boundary of $\gamma$ flip the sign of $\langle W(\gamma)\rangle$. Flips of $B_p$ in the interior of $\gamma$ do not change the sign of $\langle W(\gamma)\rangle$. However, if there happens to be a string of weak $h_e$ links going from the interior of $\gamma$ and crossing through the boundary of $\gamma$, and $B_p$ overpowers $Z_e$ along this entire string, then this would also give a perturbed disorder realization with the sign of $\langle W(\gamma)\rangle$ flipped. The probability for such a string to occur is $\left(\frac{\sqrt{2}\lambda_B}{\sqrt{\pi}\Delta_h}\right)^L$ where $L$ is the length of the string. 

If $\frac{\lambda_B}{\Delta_h}\ll 1$, then the leading contribution comes from the strings of length 1, i.e. the plaquettes at the boundary of $\gamma$. It is very unlikely for $B_p$ to overpower $\Delta_h$ on two consecutive plaquettes, so in any particular disorder realization the addition of $-\lambda_B\sum B_p$ to the Hamiltonian only leads to modification of the ground state by some pairs of nearest neighbor flips of $B_p$ (from a single $h_e$ getting overpowered). If these nearest neighbors happen to occur at the boundary of $\gamma$, then the sign of $\langle W(\gamma)\rangle$ would change. From the above reasoning, to leading order, we can ignore the bulk of $\gamma$, and only consider plaquettes at the boundary of $\gamma$. Originally, the probability to have $B_p=\pm 1$ are given by

\begin{equation}
    P(B_p=+1)=P(B_p=-1)=\frac{1}{2}
\end{equation}

with the addition of the $-\lambda_B\sum B_p$, we have
\begin{equation}
    P(B_p=+1)=\frac{1}{2}+\epsilon\qquad P(B_p=-1)=\frac{1}{2}-\epsilon
\end{equation}
where $\epsilon=\frac{\sqrt{2}\lambda_B}{\sqrt{\pi}\Delta_h}$. The change in $\langle W(\gamma)\rangle$ is given by the imbalance of $\prod_{p\in\partial\gamma}B_p$ to be $+1$ rather than $-1$. When $\lambda_B=0$, the probability for $+1$ is the same as the probability for $-1$. When $\lambda_B\neq 0$, this probability is
\begin{align}
&P\left(\prod_{p\in\partial\gamma}B_p=+1\right)\nonumber\\
&= \sum_{j=0}^{\lfloor{|\partial\gamma|/2}\rfloor} \binom{|\partial\gamma|}{2j} \left(\frac{1}{2} - \epsilon\right)^{2j}\left(\frac{1}{2}+ \epsilon\right)^{|\partial\gamma|-2j}\nonumber\\
&=\frac{1}{2} \left(1 + (2\epsilon)^{\partial\gamma}\right)
\end{align}
This means that the possibility of $B_p$ overpowering $h$ at the boundary of $\gamma$ leads to an increased probability that $\langle W(\gamma)\rangle=+1$ compared to $\langle W(\gamma)\rangle = -1$ in any particular disorder realization. It follows that $\mathrm{Tr}(\rho W(\gamma))$ decays exponentially in the perimeter of $\gamma$:
\begin{equation}
    \mathrm{Tr}(\rho W(\gamma))\sim \left(2\epsilon\right)^{|\partial \gamma|}= \left(\frac{2\sqrt{2}\lambda_B}{\sqrt{\pi}\Delta_h}\right)^{|\partial \gamma|}
\end{equation}

Note that the above makes it seem like this fails for $\sqrt{\pi}\Delta_h<2\sqrt{2}\lambda_B$, but this is only because we have expanded the error function in the limit $\lambda_B\ll \Delta_h$. If we used the exact error function rather than this approximation, then the quantity inside the parenthesis above is always $<1$, so we always get exponential decay in $|\partial\gamma|$.

\bibliography{main.bib}

\end{document}